\documentstyle[12pt]{article}
\begin{document}
\pagenumbering{roman}
\vspace*{0.5in}
\begin{center}
\noindent
{\Large\bf Statistical Mechanics of Semiflexible Chains: } \\
{\Large\bf A meanfield variational approach}     

\vspace{.25in}
{\sf D. Thirumalai} \\
{\it Institute for Physical Science and Technology\\
University of Maryland, College Park, MD 20740  }\\
 \vspace{0.3in}
{\sf B.-Y. Ha }\\
{\it Department of Chemistry and Biochemistry\\
University of California at Los Angles\\
Los Angeles, CA 90024 }\\
\vspace{1cm}
\end{center}

\newpage
\pagenumbering{arabic}
\begin{center}
{\Large\bf Statistical Mechanics of Semiflexible Chains: }\\
{\Large\bf A meanfield variational approach}  \\
\vspace{.25in}
{\sf D. Thirumalai and B.-Y. Ha} \\
\end{center}
\vspace{1in}

\noindent
{\bf 1  INTRODUCTION} \\

The random walk model for neutral polymer is perhaps the simplest mathematical 
representation for long flexible chains~\cite{1.}.  The tremendous progress made in the 
theoretical understanding of conformational and dynamical properties of 
flexible polymer systems becomes possible because
 systematic calculations using the random walk
model in conjunction with the inclusion of excluded volume 
interactions can be carried out at least in principle~\cite{2.}.  The 
resulting model, referred to as the Edwards model, is the minimal 
representation of real polymers that adequately describes the global 
properties of several polymeric systems.  The random walk model views the flexible polymer
chains as a Brownian curve.  In the discrete representation, a 
flexible chain can be modeled as one for which angles between successive
chain segments are not correlated.  Since the orientations of chain segments 
are independent, the segment vectors have the Markovian
property so that the 
mean squared end-to-end distance is proportional to the number $N$ of segments 
of size $a$ in the chain.
In the continuum limit this chain becomes
a Brownian curve.  

In contrast there are a number of issues concerning the behavior of 
semiflexible chains that are not satisfactorily solved.  Many 
polymeric molecules have internal stiffness and hence can not 
modeled as freely jointed chains~\cite{3.}.  This is especially the case in 
several biopolymers such as actin, DNA, and microtubles~\cite{Frey}.  A 
measure of the stiffness of a polymer is in terms of the so called 
persistence length, $l_{\rm p}€$, which gives an estimate of the 
length scale over which the tangent vectors along the contour of the 
chain backbone are correlated.  The typical values of $l_{\rm p}€$ of 
biopolymers lie in the range of a several nm to few mm. This range is 
several orders of magnitude larger than the persistence length of 
flexible chains.   If the contour length, $L$, of such molecules is of 
the same order of magnitude as $l_{\rm p}€$ then it is imperative to 
include bending rigidity to describe the conformations of the chain.  
In these situations the chain behaves as wormlike and the appropriate 
model for this was introduced sometime ago~\cite{3.}.  It has been known for 
a long time that the wormlike model provides a good starting point in 
the theoretical description of chains with internal stiffness.  
Inspired by several recent experiments~\cite{SFB}, which have probed a 
number of properties of stiff biological molecules, there has been a 
renewed interest in understanding their shapes.  The purpose of this 
chapter is to provide a simple way of calculating a number of 
interesting properties of semiflexible chains using a simple meanfield 
variational approach. 

The effect of excluded volume profoundly changes the properties of flexible chains.
Although historically the importance of excluded volume was recognized 
sometime ago, the introduction of the Edwards to represent the effect of this
short range interaction made possible systematic calculation of various static
and dynamic properties using field theoretical methods~\cite{2.}.  In
this chapter, in 
which we focus on the properties of semiflexible chains, the excluded volume
effects will be ignored. This is physically reasonable for relatively stiff 
chains for which deviations from rod-like conformations are negligible.

A simple way to account for the stiffness of a semiflexible chain
is to constrain the angles between two successive segments $\theta$ to be
fixed.  The value of ${\theta}$ depends on the local stiffness 
of the chain.  This prescription leads to the freely 
rotating chain model.  If we describe the configurations of a polymer chain 
by the set of position vectors 
$\{{\bf r}_{ n}\}=({\bf r}_0,...,{\bf r}_{ N})$ or 
alternatively by the set of segment vectors $\{{\Delta}{\bf r}_{ n}\}=
({\bf r}_1-{\bf r}_0,...,{\bf r}_{N}-{\bf r}_{N-1})$, then the spatial
correlation, $\bigl<{\Delta}{\bf r}_{n} \cdot {\Delta}{\bf r}_{n-1}\bigr>$, in the 
freely rotating chain has  
the assigned value $a^2 {\rm cos}\ {\theta}$.  In the continuous limit 
$(a \rightarrow 0, {\theta} \rightarrow 0, N \rightarrow \infty, Na=L)$ the 
freely rotating chain becomes the so called wormlike chain~\cite{3.}. In this case,  
the ratio $2a/{\theta}^2$ defines the {\it persistence length} $l_{\rm p}$, which is the typical length scale over which the chain 
changes its direction appreciably.  Other conformational properties of such
a model are well known in the literature [3,6-20].  

The spatial correlations 
$\bigl<{\Delta}{\bf r}_{n} \cdot {\Delta}{\bf r}_{m} \bigr>$,
which characterize the properties of a semiflexible chain, decay
exponentially as ${\rm exp}[-a \vert n-m \vert/l_{\rm p}]$.  Thus the conformational 
properties of a semiflexible chain beyond the length scale $l_{\rm p}$
reduce to those of flexible chains, i.e., one can view the stiff chain as being
made up of several rigid segments of length $l_{\rm p}$ that are freely joined.  However 
because of the intrinsic 
skeletal stiffness of many synthetic polymers as well as biopolymers one
needs to develop a model that explicitly builds effects due to chain bending.
The chain stiffness turns out to be a  
relevant parameter in the description of isotropic-nematic transition condition in 
liquid-crystalline polymers~\cite{9.}.  Even for an isolated chain, the chain stiffness 
should be taken into account to describe the local properties of stiff
polymer chains.  This is especially important in polyelectrolytes. 
The scaling behavior of the electrostatic persistence length $l_{\rm e}$ is 
known to depend on
the rigidity of the chain~\cite{HT2,BJ}.  Many biological molecules and short chains of 
otherwise flexible chains also belong to the class for which the chain stiffness
plays an important role.

A number of theoretical models have been introduced in the literature to 
account for chain stiffness.  The earliest model for stiff chains is the  
wormlike chain (also known as Kratky-Porod model) in which the angles between
successive chains are constrained~\cite{3.}.  Although physically reasonable, this model
has not yielded analytically tractable results for equilibrium and dynamical 
properties.  Harris and Hearst introduced a "simplified model" of stiff chains  
in which the tangent vector ${\bf u}(s)=\partial {\bf r}/\partial s$ was 
allowed to fluctuate as opposed to having the constraint ${\bf u}^2(s)=1$
for all $s$ ~\cite{5.}.  It has been noted that the resulting model does not
satisfy the spatial homogeneity of stiff chains.  More recently a model that does not suffer from this 
restriction was proposed by Lagowski, Noolandi, and Nickel~\cite{12.} using a 
functional integral formalism.  These authors showed that the resulting model
yielded the mean squared end-to-end distance in 
agreement with Kratky-Porod.  The spatial correlations decay exponentially
with a slightly shorter value of the persistence length.

In this chapter we show that a model for stiff chains proposed by Lagowski,  
Noolandi and Nickel (LNN)~\cite{12.} results from a stationary phase
evaluation of  
certain functional integrals that occur  in 
an appropriate field theory for stiff chains.  
Our approach is  
systematic and can be applied to many diverse problems involving semiflexible
chains.  We should
note that Winkler et al.~\cite{14.} have also obtained a model for stiff chains using the 
maximum entropy principle.  These authors did not notice that their model
in the continuum limit is identical to that of LNN.  Furthermore their method
appears more cumbersome than the standard functional integral approach presented
here. 

The rest of the chapter is organized as follows.  In section 2, we 
present the basic strategy behind the meanfield variational approach.  
The resulting model, as mentioned earlier, leads to the LNN 
representation of wormlike chains.  The theoretical ideas are used to 
calculate the distribution of end-to-end distance in semiflexible 
chains in section 3.  The application of the theory to the problem of 
semiflexible chains under tension with forces or the stretching of DNA 
is developed in section 4.  The possible limitations of the theory are 
illustrated in section 5 by studying the behavior of semiflexible 
chains in a stretching nematic field.  The chapter is concluded in 
section 6 with a few additional remarks. \\

\vspace{0.5in}
\noindent
{\bf 2  MEANFIELD MODEL} \\
\noindent
{\bf 2.1 flexible chains} \\

The basic methodology can be illustrated using the simpler example of a 
flexible chain.  This is the limiting case of a stiff chain as the rigidity 
vanishes.  The probability function for the flexible chain conformations 
without excluded volume interactions can be written as
\begin{equation}
\label{Psi{}} 
\Psi \{{\bf r}_{n}\}=\prod_{n=1}^{N} \psi(\Delta{\bf r}_{n})
\end{equation}
where $\psi = \delta(\vert \Delta{\bf r} \vert -a)/4 \pi a^2 $ denotes
the random  
distribution of a segment vector of length $a$.  Eq. (1) accounts 
for the chain connectivity.  We can rewrite
the probability 
weight in Eq. (1) by introducing auxiliary fields ${\lambda}_{n}$ as
\begin{equation}
\Psi \{{\bf r}_{n}\}
\propto \int^{i \infty}_{-i \infty} \prod_{n=1}^{N} {\rm d}{\lambda}_{n}\ {\rm exp}
\biggl[- 
\sum_{n=1}^{N} {{\lambda}_{n} \over a}
\bigl((\Delta {\bf r}_{n})^2-a^2 \bigr)\biggr]
.\end{equation}

We now show that a stationary phase evaluation of the free energy of the chain
described by above weight leads to the probability weight for the Brownian chain.
This approximation amounts to relaxing the locally enforced constraint of 
$({\Delta}{\bf r}_{n})^2=a^2$ to a global one, $\bigl<({\Delta}{\bf
r}_{n})^2 \bigr>=a^2$
,and the validity of the approximation can   
be justified {\it a posteriori} as $a \rightarrow 0$.  
  The free energy $F$ of a non-interacting flexible chain can be written as
\begin{equation}
\label{F}
{\rm exp}(- F/k_{\rm B}T) ={\rm const} \int^{i \infty}_{-i \infty} \prod_{n=1}^{N} 
{\rm d} \lambda_n 
 {\rm exp}(- {\cal F}\{ \lambda_n \})
\end{equation}
where the free energy functional ${\cal F} \{ \lambda_n \}$ is defined by   
\begin{eqnarray}
{\cal F}\{\lambda_n \} &\equiv& -{\rm ln}\Biggl\{ \int \prod_{n=1}^N {\rm d}{\bf r}_n 
 {\rm exp}\biggl[- a^{-1} \sum_{n=1}^N \lambda_n {\bf r}_n^2 \biggr] \Biggr\}
 - a \sum_{n=1}^N \lambda_n \nonumber \\
  &=& \sum_{n=1}^N \Bigl( \mbox{$\frac{3}{2}$} {\rm ln} \lambda_n -\lambda_n a \Bigr)
  +{\rm const}
.\end{eqnarray}
In the above equation, the order of the ${\bf r}_n$ and $\lambda_n$ 
integrations is interchanged.  If we 
denote the trajectory $\lambda_n$ along which the integrand in
Eq.~(\ref{F}) has its 
maximum value by $\lambda_n^{\rm cl}$, then the free energy can be expanded 
around this stationary phase trajectory, $\lambda_n^{\rm cl}$.  In the following
 the subscript `cl' will be omitted.  In the meanfield 
theory for which the constraint is imposed only on an average, we retain only 
the leading term in this expansion.  By
setting the partial derivative of the free energy functional 
${\cal F} \{ \lambda_n \}$ with respect to
$\lambda_n$ to zero, we get the stationary phase condition:
\begin{equation}
{\partial \over \partial \lambda_n} {\cal F}\{ \lambda_n \}=0 \Rightarrow \lambda_n={3 \over 2 a},
 \qquad \qquad 0 \le n \le N
.\end{equation}
The independence of ${\lambda}_{ n}$ on $n$ reflects the symmetry
of the problem of an ideal flexible chain.  Since the delta function can be 
also represented as
 \begin{equation}
\delta({\bf r})=\lim_{a \rightarrow 0} \Bigl({3 \over 2 \pi a^2} \Bigr)^{3/2}
\ {\rm exp}(-{\bf r}^2/2a^2)
,\end{equation}
 the stationary phase evaluation becomes very accurate 
in the continuum 
limit, $a \rightarrow 0$.
Thus long flexible chains, i.e., $N \gg 1$ can be well described by the 
following weight in the continuum limit
\begin{equation}
\label{Psi}
{\Psi}_{\rm MF}€[{\bf r}(s)] \propto \ {\rm exp}\biggl [- {3 \over 2a} 
\int_0^L {\rm d}s \biggl( {\partial {\bf r} \over \partial s }\biggr)^2 
\biggr]
\end{equation}
where $\Psi[{\bf r}(s)]$ is written in the functional integral notation and is 
the Wiener measure.

By treating the random fields ${\lambda}(s)$ at the mean field level, the 
microscopic constraints conjugate to the fields ${\lambda}(s)$, which 
ensure that the chain segments are connected but otherwise randomly
distributed, are relaxed to the global ones.  This results in the expected
probability weight given in Eq.~(\ref{Psi}) for a long flexible chain and is the 
Wiener measure obtained in the path integral description of a diffusion
equation. \\

\vspace{0.2in}
\noindent
{\bf 2.2  Linear stiff chains}\\

The approach described above can be extended to semiflexible chains.  In 
this
calculation we assume that the stretching of two connected chain segments 
is not important so that the coupling between this degree of freedom
and the bending degree of freedom can be ignored ~\cite{15.}.  In this
case, the weight in Eq.~(\ref{Psi{}}) needs to be modified so that it yields non-vanishing
correlations $\bigl<\Delta{\bf r}_{ n} \cdot \Delta{\bf r}_{ n-1} \bigr>= 
a^2{\theta}^2=2a^3/l_{\rm p}$.  This can be achieved if we multiply the weight
in Eq.~(\ref{Psi{}}) by the Boltzmann 
weight ${\rm exp}(l_{\rm p}a^{-3} \sum_{n=1}^{N-1} \Delta {\bf r}_{ n+1}
         \cdot \Delta {\bf r}_{ n})$ corresponding to the local 
interactions between adjacent segments.  This term favors parallel alignment
of adjacent segments over bent configurations.  In the ${\lambda}_{ n}$ 
representation of the probability weight, this can be rewritten as
${\rm exp}[-{1\over 2}l_{\rm p}a^{-3} (\Delta {\bf r}_{ n+1}- 
\Delta {\bf r}_{ n})^2]$ 
with a redefinition of ${{\lambda}}_{n}$.
Then the weight associated with 
a particular configuration of a semiflexible chain becomes
\begin{equation}
\label{Psi{}1}
\Psi \{{\bf r}_{ n}\} \propto \int^{i \infty}_{-i \infty} \prod_{n=1}^{N}{\rm d}{\lambda}_{n}
\ {\rm exp}\biggl[-\sum_{n=1}^{N}{ {\lambda}_{ n} \over a} 
((\Delta{\bf r}_{ n})^2-a^2)-{l_{\rm p}\over 2a^3} \sum_{n=1}^{N-1} 
(\Delta{\bf r}_{ n+1} 
-\Delta{\bf r}_{ n})^2 \biggr]
.\end{equation}
In the continuum limit, this can be written as functional integral
\begin{equation}
\label{Psi1}
\Psi[{\bf u}(s)] \propto\ {\rm exp}\biggl[-{l_{\rm p} \over 2} \int_0^L {\rm d}s  
\Bigl({\partial {\bf u} \over \partial s}\Bigr)^2 \biggr]\ \prod_{0\le s \le L} 
\delta ({\bf u}^2(s)-1)
\end{equation}
where ${\bf u}(s) \equiv \partial {\bf r}(s)/\partial s$ is a unit tangent 
vector.   The properties associated with the weight $\Psi[{\bf u}(s)]$ are 
well known in the literature [2-3,10].  The random variable ${\bf u}(s)$ describes the rotational
Brownian motion on a unit sphere, ${\bf u}^2=1$.  If we let 
$P({\bf u}_{ s},{\bf u}_{ s'};s',s)$ be the probability that 
${\bf u}(s')={\bf u}_{ s'}$ when ${\bf u}(s)={\bf u}_{ s}$, then 
this function obeys a diffusion equation on the unit sphere.  The solution of the 
diffusion equation can be expanded in terms of spherical harmonics.  This
enables us to compute the following correlation
\begin{equation}
\label{ecor}
\bigl<{\bf u}(s') \cdot {\bf u}(s) \bigr>={\rm exp}(-\vert s'-s \vert 
/ l_{\rm p} )
.\end{equation}
This correlation along with the Markovian property of ${\bf u}$ leads to the 
mean squared end-to-end distance given by
\begin{eqnarray}
\label{R2}
\bigl<R^2 \bigr>&=&\int_0^L \!\!\int_0^L {\rm d}s{\rm d}s' \bigl<{\bf u}(s') \cdot {\bf
u}(s) \bigr> \nonumber \\
&=&2l_{\rm p}L-2{l_{\rm p}}^2(1-{\rm e}^{-L/l_{\rm p}})
.\end{eqnarray}

Even though the results given in Eq.~(\ref{ecor}) and Eq.~(\ref{R2}) are exact, the use of 
Eq.~(\ref{Psi1}) to describe non-ideal semiflexible chains
turns out to be quite formidable.  The major difficulty arises because of the  
constraint ${\bf u}^2(s)=1$.  One encounters similar difficulty
in other physical systems described by the non-linear $\sigma$ model~\cite{16.} for which the 
magnitude of a spin ${\bf S}$ is held fixed, ${\bf S}^2={\rm const}$.  Thus it is 
of practical interest to obtain a tractable model for such constrained 
systems.  We will extend the stationary phase approach adopted for the flexible 
chain to obtain a tractable meanfield model for a semiflexible chain.  

In our stationary phase approach, the field $\lambda_{ n}$ is treated as 
a parameter to be determined.  The dependence of $\lambda_{ n}$ on $n$ 
depends on the problem
under consideration.  The free energy functional for an ideal semiflexible 
chain can be written as 
\begin{equation}
{\cal F} \{ \lambda_n \}=
-{\rm ln} \int \prod_{n=1}^{N} {\rm d}{\bf r}_{ n}        
\ {\rm exp}\biggl[-{E\over k_{\rm B}T}+
a \sum_{n=1}^{N}\lambda_{ n}  \biggr]
\end{equation}
where $E$ in a matrix form is given by
\begin{equation}
{E a \over k_{\rm B}T}={\zeta}^T Q \zeta
\end{equation}
with $\zeta \equiv \{ {\bf r}_1,...,{\bf r}_{ N}\}^T$.  The $3N \times 3N$
matrix $Q$ is defined by 
\begin{equation}
Q_{ nm}={\lambda}_{ n}{\delta}_{ n\ m}-
{l_{\rm p}\over 2a^2}(1+{\delta}_{ n \ m \pm 1})
.\end{equation} 
Then the free energy ${\cal F}$ is 
given by
\begin{equation}
{\cal F} \{\lambda_n \} ={3\over 2} \ {\rm ln}({\rm det} \ Q) -a \sum_{n=1}^{N}{\lambda}_{\rm n}
+{\rm const}
.\end{equation}
The stationary phase evaluation of ${\lambda}_{ n}$ amounts to minimizing the 
free energy with respect to ${\lambda}_{ n}$, i.e.,
\begin{equation}
\label{FM}
{\partial \over \partial \lambda_n}{\cal F} \{ \lambda_n \}=0 
\Rightarrow
{3\over 2} {\partial {\rm ln} ({\rm det}Q) \over \partial \lambda_n}=a, 
\qquad \qquad 1\le n \le N
.\end{equation}
It can be easily shown that the minimization condition in Eq.~(\ref{FM}) 
amounts to requiring $\bigl<{\bf u}^2 \bigr>=1$ in the continuous limit.  This follows 
because Eq.~(\ref{FM})  can be rewritten as   
${\partial \over \partial \lambda_n}{\cal F}
=\bigl<(\Delta {\bf r}_{ n})^2/a^2 \bigr>-1$. 
This is a simultaneous equation for the unknown parameters $\lambda_{ n}$ for 
which we can not find an analytical solution.  An examination of the 
structure of the matrix $Q$, however, leads to the following properties of 
$\lambda_{ n}$ which satisfy the above equation; $\lambda_1=\lambda_{ N}
\neq \lambda_2=,...,=\lambda_{ N-1}$.
For our
purposes it suffices
if ${\lambda}_{ n}$ can be chosen so that $\bigl<{\bf u}^2(s) \bigr>=1$ and other
conformational properties are reproduced.  
If all $\lambda_{ n}$ are
equal to each other, as is the case for the flexible chain, then the chain 
described by the probability weight in 
Eq.~(\ref{Psi{}1}) shows inhomogeneity, i.e., the chain fluctuates more strongly at both ends 
than elsewhere.  
Having recognized the translational asymmetry in the problem of a semiflexible 
chain, it is convenient to rewrite $\lambda_{ n}$ as follows; 
$\lambda_1=\lambda_{ N}=\lambda+\delta$/a, $\lambda_{ n}=\lambda \ (2\le n \le N-1)$. 
With these simplifications, the weight for the semiflexible chain at the 
level of a stationary phase approximation, becomes
\begin{equation}
\label{Psi2}
\Psi_{\rm MF}€[{\bf u}(s)] \propto {\rm exp}\biggl[-\lambda \int_0^L {\rm d}s 
{\bf u}^2(s) 
-{l_{\rm p}\over 2} \int_0^L {\rm d}s 
\Bigl({\partial {\bf u} \over \partial s}\Bigr)^2
-\delta ({\bf u}_0^2+{\bf u}_{ L}^2) \biggr]
.\end{equation}

The functional in Eq.~(\ref{Psi2}) is  identical in form to that proposed by LNN.  The 
explicit expression for det{\it Q} and thus the stationary-point conditions 
for $\lambda$ and $\delta$ can be 
obtained by setting a recursion relation in $N$.  Alternatively, we can 
exploit an analogy between the path integral in Eq.~(\ref{Psi2}) and the harmonic 
oscillator in quantum mechanics~\cite{17.}. 
 The propagator of a harmonic oscillator of a mass $l_{\rm p}$ and a
frequency $\Omega=\sqrt{ 2 \lambda/l_{\rm p}} $, denoted by
$Z({\bf u}_0,{\bf u}_{ L};L)$, is given by
\begin{equation}
Z({\bf u}_0,{\bf u}_{ L};L)=
\biggl({2 \pi \ {\rm sinh} \  \Omega L \over \Omega  l_{\rm p}} 
\biggr)^{3/2} 
 {\rm exp} \biggl[ - {({\bf u}_L^2 + {\bf u}_0^2) \ {\rm cosh} \ \Omega
L -2{\bf u}_0 \cdot {\bf u}_L \over\Omega l_{\rm p}/2 \cdot \ {\rm
sinh}  \ \Omega L  } \biggr] 
.\end{equation}
We can thus rewrite the free energy as
\begin{eqnarray}
{\cal F}[\lambda,\delta]&=&-{\rm ln} \int {\rm d}{\bf u}_0 
{\rm d}{\bf u}_{ L} \ 
{\rm e}^{-\delta ({\bf u}_0^2+{\bf u}_{ L}^2)} 
\ Z({\bf u}_0,{\bf u}_{ L};L)
-(L \lambda+2 \delta)+{\rm const} \nonumber \\
&=& \mbox{\small$\frac{3}{2}$} \ {\rm ln}\Biggl[ \Biggl(\delta \ {\rm
sinh} \ L \sqrt{2 \lambda \over l_{\rm p}}
+\sqrt{\lambda l_{\rm p} \over 2} {\rm cosh} \ L
\sqrt{2 \lambda \over l_{\rm p}} \Biggr)^2 -
{\lambda l_{\rm p} \over 2}\Biggr] \nonumber \\
&-&\mbox{\small$\frac{3}{2}$} \ {\rm ln} \sqrt{\lambda l_{\rm p} \over 2}-
 \mbox{\small$\frac{3}{2}$} \ {\rm ln}\Biggl({\rm sinh} \ L 
 \sqrt{2 \lambda \over l_{\rm p}} \Biggr)
-(L\lambda+2\delta) +{\rm const} 
\end{eqnarray}
where we have used
\begin{equation}
\int_{-\infty}^{\infty}\!\int_{-\infty}^{\infty}{\rm d}x \ {\rm d}y{\rm
e}^{-p^2(x^2+y^2) \pm q xy}={\pi
\over \sqrt{p^4-\mbox{\small$\frac{1}{4}$}q^2}} 
.\end{equation}
A little algebra leads to the following stationary phase conditions for $\lambda$
and $\delta$ which require ${\partial \over \partial 
\lambda} {\cal F}=0={\partial \over \partial \delta}{\cal F}$: 
\begin{equation}
\label{SPC}
\sqrt{\lambda l_{\rm p}\over 2} ={\delta}={3 \over 4}
\end{equation}
Note here that the values of $\lambda$ and $\delta$ don't depend on the contour
length $L$ of a chain.  Since the stationary phase condition is imposed on 
$\lambda$, only one of $l_{\rm p}$ or $\lambda$ is independent.

To understand the features implied by the weight in Eq.~(\ref{Psi2}), let us compute the 
correlation $\bigl<{\bf u}(s) \cdot {\bf u}(s') \bigr>$.  Using the Markovian property 
of ${\bf u}$, the correlation can be computed:
\begin{eqnarray}
\label{cor}
\bigl<{\bf u}(s) \cdot {\bf u}(s') \bigr>
&=&{\cal N}^{-1} \int {\cal D}[{\bf u}] \ {\bf u}(s) \cdot {\bf u}(s') \ 
{\Psi_{\rm MF}€[{\bf u}]}  \nonumber \\
&=& {\cal N}^{-1} \int {\rm d}{\bf u}_0 {\rm d}{\bf u}_s {\rm d}{\bf
u}_{s'} {\rm d}{\bf u}_L {\rm e}^{-\delta({\bf u}_0^2+{\bf
u}_L^2)}Z({\bf u}_L,{\bf u}_{s'};L-s') \nonumber \\
&{}&\times Z({\bf u}_{s'},{\bf u}_{s};s'-s) {\bf u}(s) \cdot {\bf
u}(s') Z({\bf u}_s,{\bf u}_0;s)  \nonumber \\
&=& {\partial \over \partial \alpha} \bigg|_{\alpha=0} {\rm ln}\biggl[
\int {\rm d}{\bf u}_s {\bf u}_{s'} {\rm e}^{-\delta({\bf u}_s^2+{\bf
u}_{s'}^2)+\alpha{\bf u}_s \cdot {\bf u}_{s'}}Z({\bf u}_{s'},{\bf
u}_{s};s'-s) \biggr] \nonumber \\
&=&{\rm exp}(-\vert s'-s \vert/l_0) 
\end{eqnarray}
where ${\cal N}$ is the normalization constant given by ${\cal N}=
\int {\cal D}[{\bf u}] \ {\Psi}_{\rm MF}€[{\bf u}]$ and
$l_0 \equiv {2 \over 3}l_{\rm p}$.  A direct consequence of the above correlation
is $\bigl<{\bf u}^2 \bigr>=1$ and thus the constraint ${\bf u}^2=1$ is enforced only on 
an average in the meanfield model of semiflexible chains with the weight given in
 Eq.~(\ref{Psi2}).  The above correlation can be 
compared with one in Eq.~(\ref{ecor}) obtained with the exact
weight.  A comparison of Eq.~(\ref{ecor}) and Eq.~(\ref{cor}) shows that the persistence length
for the approximate model for stiff chains (cf. Eq.~(\ref{Psi2})) is smaller by a 
factor of ${2 \over 3}$.  The plausible reason for this is the following: In the original 
model the constraint condition is ${\bf u}^2(s)=1$ for all values of $s$.  The 
model obtained by enforcing the global condition $\bigl<{\bf u}^2 \bigr>=1$ allows for 
unrestricted (restricted only on an average) fluctuations in ${\bf u}$ thus allowing for configurations that 
would be prohibited by the restricted condition ${\bf u}^2(s)=1$.  Thus we would  
expect $l_0$ to be less than $l_{\rm p}$. 

Winkler et al.~\cite{14.} have obtained 
exactly the same result (see their Eq. (4.18)) using the maximum entropy 
principle.  Since they also only enforce the constraint on an average the 
resulting theory, as described here, should be viewed as meanfield like.
If $l_0$ is understood as a new definition
of the persistence length, then the stationary phase weight in
Eq.~(\ref{Psi2}) predicts the same conformational 
behavior as the exact one in Eq.~(\ref{Psi1}).   If dimension $d$ is
large enough, 
we expect the effect of fluctuations around the mean value to become
minimal.  It is known that the stationary phase approach described here
becomes exact as $d \rightarrow \infty$ [18].   It's worth noting that for a chain
described by this weight with $\delta=0$, the above correlation holds only 
for $0 \ll s,s' \ll L$.  This is because of the excess end fluctuations in this 
case.  \\

\vspace{0.2in}
\noindent
{\bf 2.3  Closed stiff chains} \\
   
For practical purposes it's more convenient to use a translationally symmetric
model for a semiflexible chain, i.e., one described by $\Psi_{\rm MF}€[{\bf u}(s)]$ in 
Eq.~(\ref{Psi2}) with $\delta =0$.  For ring polymers for which the periodic condition ${\bf u}(0)={\bf u}(L)$ 
and the closure relation $\int^L_0 {\bf u}(s) {\rm d}s =0$ are
imposed, we expect all the ${\lambda}(s)$ to be equal or $\lambda (s)$ should
be independent of $s$.  
The free energy in this case can be written as
\begin{equation}
{\cal F}[\lambda]=-{\rm ln} \ \int_{{\bf u}_0={\bf u}_L} {\cal D}[{\bf u}(s)] 
\delta\bigl( \mbox{$\int_0^L $}{\rm d}s \ {\bf u}(s) \bigr) \Psi_{\rm MF}€[{\bf 
u}(s)]\Big|_{\delta=0}€ -L \lambda
.\end{equation}
To compute this integral, it is convenient to introduce the Fourier 
transform  
\begin{equation}
\label{FT}
{\bf u}(s)= \sum_{n=-\infty}^{\infty}{\bf u}_n{\rm exp}\biggl[i {2 \pi s n \over
L} \biggr]
.\end{equation}
Completeness of this expansion is reflected in the closure
condition 
\begin{equation}
{1 \over L} \int {\rm d}s \ {\rm exp} \bigg[ i{2 \pi s  \over L}(n'-n)
\bigg]=\delta_{nn'}
.\end{equation}
In terms of new variables ${\bf u}_n$, the weight can be 
written as
\begin{equation}
\label{Psi{}2}
\Psi_{\rm MF}€\{{\bf u}_n \}={\rm exp} \biggl\{- 
 \sum_{n \ne 0} \biggl[{l_{\rm p} \over 2}
{(2 \pi n )^2 \over L}+\lambda L  \biggr] {\bf u}_n \cdot {\bf u}_{-n}  
\biggr\}
\end{equation}
where the contribution from $n=0$ is excluded to incorporate the
closure condition ${\int_0^L {\bf u}(s) {\rm d}s=0}$.
Consequently free energy becomes
\begin{equation}
{\cal F}[\lambda]=-{\rm ln} \int \prod_{n \ne 0} {\rm d}{\bf u}_n 
\Big| {\rm det} \mbox{$\frac{\partial {\bf u}(s)}{\partial {\bf u}_n} $} 
\Big|
\ \Psi_{\rm MF}€ \{{\bf u}_n \} -L \lambda
 .\end{equation}
 Since the transformation in Eq.~(\ref{FT}) linear, the Jacobi determinant 
 $\Big| {\rm det} \mbox{$\frac{\partial {\bf u}(s)}{\partial {\bf u}_n} $} 
 \Big|$ is 
 independent of ${\bf u}_n$.  Furthermore it does not depend on $\lambda$.   
Its value is thus unimportant and will give rise to an additive constant to  
the free energy:
\begin{eqnarray}
{\cal F}[\lambda]&=&-{\rm ln} \ \int \prod {\rm d}{\bf u}_n \Psi_{\rm 
MF}€\{ {\bf u}_n \}
-L \lambda + {\rm const.}\nonumber \\
&=& 
\mbox{\small$\frac{3}{2}$} \sum_{n \ne 0} {\rm
ln} \bigg(1+ {\lambda L^2 \over 2 l_{\rm p} n^2 \pi^2} \biggr) +
{\rm const.} \nonumber \\
&=&3 \ {\rm ln} \Biggl({\rm sinh}L \sqrt{\lambda \over 2 l_{\rm p}} \Biggr)
-\mbox{\small$\frac{3}{2}$} \ {\rm ln}(L \lambda)-L \lambda +{\rm const.}
\end{eqnarray}
where const. refers to $\lambda$-independent terms.  In the last step
of the above equation, we have used
\begin{equation}
{\rm ln} \ {\rm sinh} \ x={\rm ln} \ x +\sum_{n=1}^{\infty} {\rm
ln} \biggl(1+{x^2 \over n^2 \pi^2} \biggr) 
.\end{equation}
Now the stationary phase condition reads 
\begin{equation}
\sqrt{l_{\rm p} \lambda \over 2}=\mbox{\small$\frac{3}{4}$} \Biggl(
{\rm coth}L \sqrt{\lambda \over 2 l_{\rm p}}-{1 \over L} 
\sqrt{2l_{\rm p} \over \lambda}  \Biggr)
.\end{equation}
Here we have $L$-dependent condition for $\lambda$.  This is because only
paths which satisfy the cyclic conditions, ${\bf u}(L)={\bf
u}(0)$ and $\int_0^L {\bf u}(s){\rm d}s$,
contribute to the free energy ${\cal F}$.
The cyclic conditions are  also incorporated in the correlation
of ${\bf u}(s)$, i.e., 
\begin{eqnarray}
\label{cor3}
\bigl<{\bf u}(s') \cdot {\bf u}(s) \bigr>&=& 
\sum_{n,n' \ne 0} \bigl<{\bf u}_n \cdot {\bf u}_{n'} \bigr> {\rm exp} \biggl[ i {2
\pi \over L}(s'n'+sn) \biggr] 
\nonumber \\
&=&\mbox{$\frac{3}{2}$} \sum_{n \ne 0} { {\rm exp}\Bigl[i {2
\pi n \over L}(s'-s) \Bigr] \over  \lambda L+ {l_{\rm p} \over 2}{(2 \pi
n)^2 \over L}  } 
\end{eqnarray}
where $\bigl<...\bigr>$ is defined by
\begin{equation}
\bigl<{\bf u}_n \cdot {\bf u}_{n'} \bigr>={\int \prod_{n \ne 0} {\rm d}{\bf u}_n {\bf u}_n
\cdot {\bf u}_{n'}  
\Psi_{\rm MF}€ \{ {\bf u}_n \}  \over \int \prod_{n \ne 0} {\rm d}{\bf u}_n
\Psi_{\rm MF}€ \{ {\bf u}_n \} }
.\end{equation}
Upon using the following identity
\begin{equation}
\label{ide}
\sum_{n=1}^{\infty} { {\rm cos} \ nx \over n^2 + \alpha^2}={\pi \over
2 \alpha} \cdot { {\rm cosh} \ \alpha (\pi- x) \over {\rm sinh} \ \alpha
\pi}-{1 \over 2 \alpha^2}
\end{equation}
we can rewrite the correlation of the unit tangent vectors in a closed form
\begin{equation}
\bigl<{\bf u}(s') \cdot {\bf u}(s) \bigr>=
{{\rm cosh}[\Omega(L-2 \vert s'-s \vert) /2]-2/\Omega L 
                    \cdot {\rm sinh}(\Omega L/2)
                                \over {\rm cosh}(\Omega L/2) 
                                -2/\Omega L \cdot {\rm sinh}(\Omega L/2)}
.\end{equation} 
Note here that $\bigl<{\bf u}(s+L) \cdot {\bf u}(s) \bigr>=\bigl<{\bf
u}^2(s) \bigr>=1$.  In the 
limit of $L \rightarrow \infty$, however, the cyclic conditions are irrelevant.
That is the stationary phase 
condition and the correlation given above reduce to those of open chains.
This can be checked by taking the limit $L \rightarrow \infty$ in above equations
for the ring polymers.  
An alternative method (without derivation) for ring segments has been proposed
 earlier~\cite{19.}.  This involves modifying the original Harris-Hearst model for open 
 chains.  Huber et al. have computed quasi-elastic scattering for a modified
 version of this model~\cite{20.}. \\

\vspace{0.2in}
\noindent
{\bf 2.4  Linear periodic chains}\\

The ring polymer described by Eq.~(\ref{Psi{}2}) was introduced to circumvent
difficulties associated with 
inhomogeneity in the linear chain with a uniform stationary phase value
$\lambda$.  According to the correlation function in 
Eq.~(\ref{cor3}), however,  we can assume a uniform $\lambda$ even for the
chains with periodic boundary conditions only without violating
the homogeneity of the chain.  In this case we should include $n=0$
contribution in the Fourier modes; the free energy is thus given by
\begin{equation}
{\cal F}[\lambda]=3 \ {\rm ln} \Biggl({\rm sinh}L 
\sqrt{\lambda \over 2 l_{\rm p}} \Biggr)
-L \lambda +{\rm const}
.\end{equation}
This leads to the following stationary phase condition
\begin{equation}
\sqrt{l_{\rm p} \lambda \over 2}=\mbox{\small$\frac{3}{4}$} \
{\rm coth}L \sqrt{\lambda \over 2 l_{\rm p}} 
.\end{equation}
The periodic boundary condition is also incorporated in the
correlation function which reads
\begin{equation}
\bigl<{\bf u}(s) \cdot {\bf u}(s') \bigr>= {{\rm cosh}[(L-2 \vert s' -s \vert) \Omega/2]
                               \over {\rm cosh}(\Omega L/2)}
.\end{equation}
This correlation shows that the chain with periodic boundary condition is 
homogeneous; $\bigl<{\bf u}^2(s) \bigr>=1 $ for all $s,\ 0 \le s \le L$.
As $L \rightarrow \infty$, the periodic boundary condition is irrelevant as
in the ring polymers.

An alternative and useful presentation of this calculation can be made 
writing the curvature term in a symmetric way.  Due to the periodic
boundary conditions, i.e., ${\bf u}(0)={\bf u}(L)$ and 
${\partial \over \partial s}{\bf u}(0) ={\partial \over
\partial s } {\bf u}(L)$,
the free energy functional for the semiflexible chain can be written as
\begin{equation}
\label{calF}
{\cal F}[\lambda]=-{\rm ln} \ \int {\cal D}{\bf u}(s) {\rm exp} \biggl[
 -\mbox{\small$\frac{1}{2}$} \int_0^L\!\int_0^L {\rm d}s {\rm d}s' {\bf u}(s)
 Q(s,s'){\bf u}(s) \biggr] -\int {\rm d}s \lambda (s)
\end{equation}
where the operator $Q$ is defined by   
\begin{equation}
\label{Q}
Q(s,s')=\biggl[-l_{\rm p} \biggl({\partial \over \partial s}\biggr)^2+2 
\lambda(s) \biggr]
\delta(s'-s) 
.\end{equation}
Integration with respect to ${\bf u}(s)$ leads to 
\begin{equation}
{\cal F}[\lambda]=\mbox{\small$\frac{3}{2}$}\ {\rm tr} \ {\rm ln} \
Q-\int {\rm d}s \lambda(s)
.\end{equation}
The stationary phase condition can be obtained by requiring $\partial
{\cal F}/\partial \lambda(s)=0$:
\begin{equation}
1=\mbox{$\frac{3}{2}$} \Bigl( {2 \over Q} \Bigr)_{s,s}=\mbox{$\frac{3}{2}$}
\Biggl( {1 \over -{l_{\rm p} \over 
2}{\partial^2 \over \partial s'^2}+\lambda(s')} \Biggr)_{s,s}
.\end{equation}
If a uniform stationary phase value $\lambda$ is assumed, the above
equation can be Fourier transformed.  To this end, it is convenient to
define a complete set of eigenstates $\{|s \big>\}$ with $s$ a
curvilinear space 
label, and a complete set $\{|n \big>\}$, Fourier-conjugate to this, such 
that ${\partial \over \partial s}|n \big> =i{2
\pi n \over L} |n \big>$ .  Thus $|s \big>$ and $|n \big>$ are related with
each other through $\big< n|s \big>={1 \over \sqrt{L}}{\rm exp}[i{2 \pi n s
\over L}]$.  With the aid of these, the right hand side of the above equation can
be written as
\begin{eqnarray}
\mbox{$\frac{3}{2}$} \Biggl( {1 \over -{l_{\rm p} \over 
2}{\partial^2 \over \partial s'^2}+\lambda} \Biggr)_{s,s}
&=&\mbox{$\frac{3}{2}$}\big<s|\Biggl( {1 \over -{l_{\rm p} \over 
2}{\partial^2 \over \partial s'^2}+\lambda} \Biggr) |s \big>
\nonumber \\
&=&\mbox{$\frac{3}{2}$}\sum_{n=-\infty}^{\infty}\big<s|\Biggl( {1
\over -{l_{\rm p} \over  
2}{\partial^2 \over \partial s'^2}+\lambda} \Biggr) |n \big>
\big<n|s \big>  \nonumber \\
&=&\mbox{$\frac{3}{2}$} \ \sum_{n=-\infty}^{\infty} {1 \over \lambda L
+ {l_{\rm 
p} \over 2}{(2 \pi n)^2 \over L}} 
.\end{eqnarray}
Combined with the identity in Eq.~(\ref{ide}), the stationary phase condition
relation can be rewritten as
\begin{equation}
\sqrt{l_{\rm p} \lambda \over 2} =\mbox{\small$\frac{3}{4}$} \
{\rm coth}L \sqrt{\lambda \over 2 l_{\rm p}} 
.\end{equation}
This condition becomes the same one  as in Eq.~(\ref{SPC}).
The model described by Eq. (35) with $L \rightarrow \infty$
leads to the same result 
for the stationary phase condition and correlation of ${\bf u}(s)$ as
the previous one in Eq.~(\ref{Psi2}).  In  
section 4, we adopt this approach to examine the elastic response of a
semiflexible chain under tension. \\

\vspace{0.2in}
\noindent
{\bf 2.5  Operator representations} \\

For periodic chains, we can exploit the analogy with quantum 
mechanics and use an operator representation of the free energy.
This is especially useful as $L \rightarrow \infty$ which allows  
the use of ground state dominance approximation.
If we interpret $\hat{\bf p} \equiv l_{\rm p} {\partial \over \partial 
s} {\bf u} $ as 
the momentum operator and 
$\hat{\bf u}$ as the position operator  
such that they satisfy the commutation relation $[\hat p_j, \hat u_k]=-i \delta_{jk}$, the free energy of  
periodic chain can be written as~\cite{18.} 
\begin{equation}
{\cal F}[\lambda]=-{\rm ln} \ {\rm tr} \ {\rm e}^{-L \hat{\cal H}[\lambda]}-L \lambda
\end{equation}
where the Hamiltonian $\hat{\cal H}$ is defined by
\begin{equation}
\label{hamiltonian}
\hat{\cal H}[\lambda]={\hat{\bf p}^2 \over 2 l_{\rm p}}+ \lambda \hat{\bf u}^2
.\end{equation}
This problem is equivalent to a 3-dimensional quantum harmonic oscillator
moving in imaginary time $s=-it$; if we denote the energy eigenvalues of the 
Hamiltonian in Eq.~(\ref{hamiltonian}) by $E_{\bf n}$ with 
${\bf n}=(n_1,n_2,n_3)$ and $n_i=0,1,2...$, then  
\begin{eqnarray}
{\rm tr} \ {\rm e}^{-L \hat{\cal H}[\lambda]}
&=& \sum_{n_1,n_2,n_3=0}^{\infty}{\rm e}^{-L E_{\bf n}}  \nonumber \\
&=& \biggr( \sum_0^{\infty} {\rm e}^{-L \Omega (n+\mbox{$\frac{1}{2}$}) }
\biggr)^3 \nonumber \\
&=&\biggl( {1 \over 2 {\rm sinh}(\mbox{$\frac{1}{2}$} \Omega L)} \biggr)^3
.\end{eqnarray}
This calculation gives
the same stationary phase condition as before. 

This approach is especially useful as $L \rightarrow \infty$.  For large $L$, we can use the 
following formula~\cite{18.}
\begin{equation}
{\rm e}^{-L \hat{\cal H}}={\rm e}^{-L E_0} \Bigl[ {\big {\vert}} 0 \big>\big< 
0 {\big {\vert}} + {\cal O}
                 \Bigl( {\rm e}^{-L(E_1-E_0)} \Bigr) \Bigr]
\end{equation}
where $E_1={3 \over 2} \Omega$ and $E_1={5 \over 2} \Omega$.  The ground state denoted by 
${\big {\vert}} 0 \big>$ corresponds to the lowest eigenvalue $E_0$ of 
$\hat{\cal H}$ and is assumed
to be unique and separated by a gap from $E_{1}€$.
As $L \rightarrow \infty $, this expression is dominated by the ground state. 
The term tr exp$(-L \hat{\cal H})$ can be easily computed to yield the stationary phase 
condition, i.e., $\sqrt{\lambda l_{\rm p} \over 2}={3 \over 4}$.
Introducing $\hat{\bf U}(s)$, the interaction representation of the operator
$\hat{\bf u}$,
\begin{equation}
\hat{\bf U}(s)={\rm e}^{-s \hat{\cal H}} \ \hat{\bf u} \ {\rm e}^{s \hat{\cal H}}
\end{equation}
we can express the correlation of ${\bf u}$ as 
\begin{equation}
\bigl< {\bf u}(s) \cdot {\bf u}(s') \bigr>={{\rm tr}[\hat{\bf U}(s)
\cdot \hat{\bf U}(s') c 
{\rm e}^{-L \hat{\cal H}}] \over {\rm tr} \
{\rm e}^{-L \hat{\cal H}}}, \qquad \qquad 
0 \le s \le s' \le L
.\end{equation}
As $L \rightarrow \infty$, the ground state is dominant in the above expression
contribution:
\begin{equation}
\bigl<{\bf u}(s) \cdot {\bf u}(s') \bigr>=\big<0 {\big \vert} {\cal T} \hat{\bf U}(s) \cdot \hat{\bf U}(s') 
{\big \vert} 0 \big>
\end{equation}
where ${\cal T}$ is a time ordering operator. In general, the correlation 
functions of the statistical analog correspond with
the time-ordered products of the corresponding quantum fields.  If we use 
the 
basis $ \vert {\bf n} \big>$ in which $\hat{\cal H}$ is diagonal, the correlation can be further simplified 
to yield
\begin{eqnarray}
\label{cor1}
\bigl<{\bf u}(s') \cdot {\bf u}(s) \bigr>&=&
 \big<0 {\big \vert} \hat{\bf u} {\rm e}^{-\vert s' - s \vert
\hat{\cal H}} \ \hat{\bf u} \  
 {\rm e}^{\vert s' - s \vert \hat{\cal H}}  {\big \vert} 0 \big> \nonumber \\
&=& \sum_{{\bf n}} {\big \vert}  \big<0 {\big \vert} \hat{\bf u} {\big \vert} 
{\bf n} \big> {\big \vert}^2 
{\rm e}^{-(E_{\bf n}-E_0) \vert s'-s \vert}  \nonumber \\
&=&{\rm e}^{-(E_1-E_0)\vert s'-s \vert}   \nonumber \\
&=&{\rm e}^{-\vert s'- s \vert/l_0}
.\end{eqnarray} 
As we have seen, the operator representation of the free energy are especially
useful for a very long chain.  In general, we can make ground state dominance 
approximation for a long chain as long as the Hamiltonian of the statistical
analog $\hat{\cal H}$ is hermitian bounded 
below and has a discrete energy spectrum.  In this case, the free energy per
unit length is approximately equal to the ground state energy of $\hat{\cal H}$.
The calculation of the correlation function reduces to calculation of transition
amplitudes $\bigl<0 \vert \hat{\bf u} \vert {\bf n} \bigr>$.  For an ideal semiflexible
chain, $\hat{\bf u}$ connects $\big<0 \vert$ with the first excited state with 
non-vanishing transition amplitude, resulting in the 
correlation function given in Eq.~(\ref{cor1}). \\

\vspace{0.5in}
\noindent
{\bf 3  END-TO-END DISTRIBUTION FUNCTION}\\

There have been several studies over the years which have focused on 
the distribution of end-to-end distance of semiflexible chains.  Most of these studies have 
started by considering the chains near the rod limit and have computed 
corrections in powers of $t^{-1}$ where $t$ is the ratio of the bare 
persistence length to the contour length.  In general the calculations 
have been done only to low orders in $t^{-1}€$.  These calculations 
are quite involved, and more importantly, they do not provide reliable 
results in the interesting cases, i.e., when $t$ is on the order of 
unity.  In a very recent paper Wilhelm and Frey~\cite{W.F} have reported 
analytic and numerical (Monte Carlo) calculations for the radial 
distribution function for a range of values $t$.  The analytical 
expressions are given in terms of an infinite series involving the 
second Hermite polynomials for chains in three dimensions.  The 
meanfield like theory presented in the previous section can be used to 
derive a very simple expression for the distribution function of 
end-to-end distance.

The distribution function of the end-to-end distance $R$ is 
\begin{equation}
G(R;L)=\biggl< \delta( {\bf R}-\int_{0}^{L}€{\bf u}(s){\rm d}s ) 
\biggr>
\end{equation}
where the average is evaluated as 
\begin{equation}
\bigl<\ldots \bigr>={ \int {\cal D}[{\bf u}(s)]\ldots \Psi_{\rm MF}€[{\bf u}(s)] 
  \over \int {\cal D}[{\bf u}(s)] \ \Psi_{\rm MF}€[{\bf u}(s) ]}
.\end{equation}
The weight $\Psi[{\bf u}(s)]$ is (see Eq.~(\ref{Psi1}))
\begin{equation}
\label{PsiG}
\Psi[{\bf u}(s)] \propto\ {\rm exp}\biggl[-{l_{\rm p} \over 2} \int_0^L {\rm d}s  
\Bigl({\partial {\bf u} \over \partial s}\Bigr)^2 \biggr]\ \prod_{0\le s \le L} 
\delta ({\bf u}^2(s)-1)
.\end{equation}
Following the previous section, at the level of the stationary 
phase approximation, the weight can be replaced by
\begin{equation}
\label{PsiMF} 
\Psi_{\rm MF}€[{\bf u}(s)] \propto {\rm exp}\biggl[
-{l_{\rm p}\over 2} \int_0^L {\rm d}s 
\Bigl({\partial {\bf u} \over \partial s}\Bigr)^2
-\lambda \int_0^L {\rm d}s {\bf u}^2(s) 
-\delta ({\bf u}_0^2+{\bf u}_{ L}^2) \biggr]
.\end{equation}
where ${\bf u}_{0}€={\bf u}(0)$ and ${\bf u}_{L}€={\bf u}(L)$.  The 
parameters $\lambda$ and $\delta$, which are used to enforce the 
constraint ${\bf u}^{2}€(s)=1$ by a global constraint $\bigl<{\bf 
u}^{2}(s) \bigr>=1$, are determined variationally.  From the result in 
sec. 2.2, it is clear that due to chain fluctuations at the ends the 
global constraint $\bigl< {\bf u}^{2}€(s) \bigr>$ cannot be imposed 
using only one parameter as suggested by others~\cite{otto}.  The variational 
solutions for $\lambda$ and $\delta$ depend on the problem under 
consideration.  For the calculation of $G(R;L)$ the optimal values of 
$\lambda$ and $\delta$ are different from that obtained in section 2.

We calculate $G(R;L)$ by replacing the true weight in Eq.~(\ref{PsiG}) by 
$\Psi_{\rm MF}€[{\bf u}(s)]$ given in Eq.~(\ref{PsiMF}).  The equation for 
$G(R;L)$ with the weight given by $\Psi_{\rm MF}€[{\bf u}(s)]$ is 
\begin{equation}
\label{G}
G(R;L)={\cal N}^{-1}€ \int_{-i \infty}^{i \infty} {{\rm d}{\bf k} 
\over (2 \pi)^{3}} \int {\rm d} \lambda {\rm d} \delta \int {\cal 
D}[{\bf u}(s)] \ {\rm e}^{-{\bf k} \cdot({\bf R}-\int_{0}^{L}{\rm d}s {\bf u}(s))}€ \
\Psi_{\rm MF}€[{\bf u}(s)]
\end{equation}
where ${\cal N}$ is an appropriate normalization constant.  The 
functional integral over ${\bf u}(s)$ in Eq.~(\ref{G}) is done by 
replacing ${\bf u}(s)$ by ${\bf u}-{{\bf k} \over 2 \lambda}$.  The resulting 
path integral corresponds to a harmonic oscillator that makes a 
transition from ${\bf u}_{0}€-{{\bf k} \over 2 \lambda}$ to ${\bf 
u}_{L}€-{{\bf k} \over 2 \lambda}$ in ``time'' $L$.  Using the standard result 
for harmonic oscillator propagator the distribution function $G(R;L)$ 
becomes
\begin{equation}
\label{G1}
G(R;L)=\int {\rm d} \lambda \int {\rm d} \delta \ {\rm exp} \Bigl\{ -{\cal 
F}[\lambda,\delta] \Bigr\}
\end{equation}
where
\begin{equation}
{\cal F}[\lambda,\delta]= \mbox{$\frac{3}{2}$} \Biggl[{\rm ln} 
\biggl({L \lambda+2 \delta \over 4 \lambda^{2}€} \biggr)+{\rm ln} 
\biggl( {{\rm sinh} \Omega L \over \Omega L} \biggr) +{\rm ln}(\alpha 
\beta) \Biggr]- \lambda L -2 \delta + { R^2 \lambda^2 \over \lambda L + 
2 \delta}+{\alpha \over \beta}
,\end{equation}
with
\begin{equation}
\alpha=\delta +l_{\rm p}€ \Omega \ {\rm coth} \biggl({\Omega L \over 2} 
\biggr)
,\end{equation}
\begin{equation}
\beta=\delta + {l_{\rm p}€ \Omega \over 2} {\rm tanh}\biggl({ \Omega L \over 
2 }  \biggr)+{1 \over \lambda L + 2 \delta}
,\end{equation}
and
\begin{equation}
\Omega=\sqrt{ 2 \lambda \over l_{\rm p} }
.\end{equation}
We evaluate the integrals over $\lambda$ and $\delta$ (see 
Eq.~(\ref{G1})) 
by the stationary phase approximation.  This approximation replaces 
the constraint ${\bf u}^{2}€(s)=1$ by a global constraint $\bigl< 
{\bf u}^{2}€(s) \bigr>=1$.  If the global constraint is enforced using 
only one variational parameter, as suggested elsewhere~\cite{otto}, then one 
simply obtain the incorrect Gaussian expression for $G(R;L)$.  The 
presence of the parameter $\delta$ accounts for the suppression of 
the fluctuations of the ends of the chain.  The stationarity condition
\begin{equation}
{ \partial \over \partial \lambda} {\cal F}[\lambda,\delta] =0
\end{equation}
gives, after some algebra,
\begin{equation}
\sqrt{{\lambda l_{\rm p}€ \over 2}}=\mbox{$\frac{3}{4}$} \biggl({1 
\over 1-{R^2 \over L^2 }} \biggr)
\end{equation}
and we find that $\delta$ does not show any significant variation 
with $R$.  Using the stationary values of $\lambda$ and $\delta$ the 
distribution of the end-to-end distance for the semiflexible chain 
becomes
\begin{equation}
\label{G2}
G(R;L)={\rm const}{1  \over (1-{R^{2}€ \over L^{2}€})^{9/2} } \ 
{\rm exp} \biggl[ -{ 9 L \over 8 l_{\rm p} } {1 \over 
(1-{R^{2}€ \over L^{2}€} ) }\biggr]
.\end{equation}
We showed in the previous section that within the stationary phase 
approach, which enforces thae constraint only globally, the 
persistence length is reduced to $l_{0}={2 \over 3} l_{\rm p}$ 
instead of $l_{p}€$.  If we let $t={L \over l_{0}€}$ the radial 
probability density in three dimensions for semiflexible chains is 
given by the simple expression, namely,
\begin{equation}
P(r;t)=4 \pi {\cal C} {r^{2}€ \over (1-r^{2}€)^{9 / 2}€} {\rm exp} \biggl[
-{3 t \over 4} {1 \over (1-r^{2}€)} \biggr]
\end{equation}
where $r={R \over L}$.  The normalization constant ${\cal C}$ which is 
determined using the condition
\begin{equation}
\int_{0}^{1}€ {\rm d} r P(r;t)=1
\end{equation}
and is given by
\begin{equation}
{\cal C}={1 \over \pi^{3/2}€ {\rm e}^{-\alpha}€ \alpha^{-3/2}€ (1+3 
\alpha^{-1}€+\mbox{$\frac{15}{4}$} \alpha^{-2}€)}
\end{equation}
where $\alpha=3t/4$.

The distribution function, $P(r;t)$, goes as $r^{2}€$ as $r 
\rightarrow 0$ and vanishes at $r=1$.  The peak of the distribution 
function occurs at
\begin{equation}
r_{\rm max}€=\sqrt{\eta + \sqrt{\eta^{2}€+14} \over 7} 
\end{equation}
where $\eta={5 \over 2}-{3 \over 4}t$.  In Fig. (1) we plot $P(r;t)$ 
for the five values of $t$ for which Wilhelm and Frey~\cite{W.F} have 
presented simulation data.  We find that our simple expression in 
Eq.~(\ref{G2}) almost quantitatively reproduces the data with the maximum 
deviation of about 10${\%}$ at the peak for $t=0.5$.  

We have also calculated the first two moments of the distribution 
function and find that they reproduce the exact results in both the 
random coil limit and the rod limit.  The distribution function 
$P(r;t)$ also has the correct limiting behavior $( \sim 
\delta(1-r))$ as $t \rightarrow 0$.  Thus the meanfield variational 
approach yields a simple expression for $P(r;t)$ that is in 
quantitative agreement with simulation and hence can form the basis 
to analyze experiments. \\

\vspace{0.5in}
\noindent
{\bf 4  SEMIFLEXIBLE CHAINS UNDER TENSION} \\

The meanfield theory described in the previous section is especially useful
when one encounters semiflexible chains in the presence of fields for which 
tractable calculations are difficult.  In this section we use the 
variational theory to investigate the external field on the conformational properties
of a semiflexible chain.  Even though the analogous problems in flexible chains
was extensively studied,  the corresponding problems in the semiflexible 
chains has only recently attracted much attention [27-32].  The 
earliest theoretical paper dealing with this problem was initiated by 
Fixman and Kovacs~\cite{F.K}.  These authors used a modified version 
of the Gaussian model for stiff chains and provided expressions for 
the stretching as a function of applied force.  Their treatment is 
only valid when the applied force is small and significant deviation 
from these predictions are observed at sufficiently large values of 
the external force.  Marko and Siggia [31-32] have calculated the 
extension as a function of force for wormlike chains and found that 
their results fit the experimental data very well.  Some aspects of 
this theory have also been considered by Odijk~\cite{21.} who also discusses the 
competition between entropically dominated effects and elasticity 
effects.  The meanfield type approach
adopted here is, we believe, more general than those 
adopted in the literature.

Gaussian chains can be arbitrarily extended under tension  $f$ as implied by
 Hook's law of elastic response, i.e., $R \sim Llf $; real chains
 cannot be extended beyond the contour length $L$. As the 
 magnitude of tension exerted on both ends of a semiflexible chain increases  
 one expects an interesting crossover to occur from Hook's limit to 
fully extended limit ($R=L$).  Thus the problem of stretched 
semiflexible chains entails a competition between entropy dominated 
effects and ordering due to the external field ${\bf f}$.  We will 
show that the meanfield treatment quantitatively reproduces 
experiments on DNA~\cite{SFB} thus providing further evidence for the validity of meanfield
 approach.  \\

\vspace{0.2in}
\noindent
{\bf 4.1  Stationary phase condition} \\

For mathematical convenience, we will adopt translationally symmetric chain model 
introduced in Eq.~(\ref{calF}).  Each segment in a semiflexible chain is assumed 
to be stretched by tangential force ${\bf f}(s)$ which tends to suppress
chain fluctuations.   The weight for the case of semiflexible chains 
in the presence of the stretching is 
\begin{equation}
\label{Psi.Ela}
\Psi [{\bf u}(s),\lambda(s)] \propto {\rm exp} \biggl[
-\mbox{$\frac{1}{2}$} \int_0^L\!\int_0^L {\rm d}s {\rm d}s' {\bf u}(s) Q(s,s') {\bf
u}(s') + \int_0^L  {\bf u}(s) \cdot {\bf f}(s){\rm d}s \biggr]
\end{equation}
where $Q$ is the same operator as introduced in Eq.~(\ref{Q}), i.e.,
\begin{equation}
Q(s,s')=\biggl[-l_{\rm p} \Bigl({\partial \over \partial s}\Bigr)^2+2 \lambda
\biggr] \ \delta(s'-s)
.\end{equation}
The last term in the exponent
 is the energy penalty for chain
conformations which are not parallel to ${\bf f}(s)$.  
The free energy can be written as
\begin{eqnarray}
{\rm exp}(-F/k_{\rm B}T) &\propto& \int_{-i \infty}^{i \infty} 
{\cal D}[\lambda(s)] \int{\cal D}[{\bf u}(s)] \  
{\rm e}^{\int_0^L
\lambda(s) {\rm d}s } \ 
\Psi[{\bf u}(s),\lambda(s)] \ 
\nonumber \\
&\propto& \int_{-i \infty}^{i \infty}{\cal D}[\lambda(s)] 
  {\rm exp}\Bigl\{-{\cal F}[\lambda(s),{\bf f}(s)] \Bigr\} 
\end{eqnarray}
where ${\cal F}$ is the free energy functional given by
\begin{equation}
{\cal F}[\lambda(s),{\bf f}(s)]=-{\rm ln} \int {\cal D}[{\bf u}(s)] 
\Psi[{\bf u}(s),\lambda(s)] 
 -\int_0^L \lambda(s) {\rm
d}s+{\rm const.}
.\end{equation}
Integration with respect to ${\bf u}(s)$ yields 
\begin{eqnarray}
\label{free.f}
{\cal F}[\lambda(s),{\bf h}(s)]&=&\mbox{$\frac{3}{2}$} \ {\rm tr} \ {\rm ln} \ Q - \mbox{$\frac{1}{2}$}
\int_0^L\!\int_0^L {\bf f}(s) Q^{-1}(s,s') {\bf f}(s') \nonumber \\
&{}&- \int_0^L \lambda(s) {\rm d}s+
{\rm const.}
.\end{eqnarray}
Note that ${\cal F}[\lambda(s),{\bf f}(s)]$ is the generating
functional for the connected correlation function defined by
\begin{eqnarray}
\label{cc}
\bigl< {\bf u}(s) \cdot {\bf u}(s') \bigr>_{\rm c}&=&
-{\delta \over \delta {\bf f}(s)} \cdot {\delta {\cal
F} \over \delta {\bf f}(s')} \nonumber \\
&=&{\partial \over \delta {\bf f}(s)} \cdot \bigl< {\bf u}(s')
\bigr> \nonumber \\
&=& \bigl< {\bf u}(s) \cdot {\bf u}(s') \bigr>-
\bigl<{\bf u}(s) \bigr> \cdot \bigl< {\bf u}(s') \bigr> 
.\end{eqnarray}
As ${\bf f}$ goes to zero, $\bigl<{\bf u}(s) \bigr>$ vanishes.  In this case, the
distinction between the connected correlation function and the
correlation function disappears.  
Performing the functional
differentiations in Eq. (46), we get
\begin{equation}
\bigl< {\bf u}(s) \cdot {\bf u}(s') \bigr>_{\rm c}=3 Q^{-1}(s,s')
\end{equation}
and
\begin{equation}
\bigl<{\bf u}(s) \bigr>=-{ \delta {\cal F} \over \delta {\bf f}(s)}
=\int_0^L Q^{-1}(s,s') {\bf f}(s')
.\end{equation}
Integrations of each term in Eq.~(\ref{cc}) with respect to $s$ and $s'$ lead to
the following 
expression for $\Delta R^2$
\begin{eqnarray}
\label{DeltaR^2}
\Delta R^2 &=&\bigl< R^2 \bigr>-\bigl<R \bigr>^2 \nonumber \\
&=& \int_0^L\!\int_0^L \bigl< {\bf u}(s) \cdot {\bf u}(s') \bigr> {\rm
d}s {\rm d}s' -\int_0^L\!\int_0^L
\bigl<{\bf u}(s) \bigr> \cdot \bigl< {\bf u}(s') \bigr> {\rm d}s {\rm
d}s' \nonumber \\
&=&\int_0^L\!\int_0^L {\delta \over \delta {\bf f}(s)} \cdot {\delta {\cal
F} \over \delta {\bf f}(s')} {\rm d}s {\rm d} s'
.\end{eqnarray}
All relevant quantities can be thus expressed in terms of the free
energy functional ${\cal F}[\lambda(s),{\bf f}(s)]$.

Following the general formalism described in section 2, the case of ideal 
semiflexible chains, the free energy is
approximated by the minimum value of the corresponding free energy
functional which occurs along the stationary phase trajectory
$\lambda$.  The stationary
phase condition is obtained by requiring ${\delta \over
\delta \lambda(s)}{\cal F} =0$.  To this end, we first note that
\begin{eqnarray}
{\delta \over \delta \lambda(s)} {\rm tr} \ {\rm ln} \ Q &=&
{\delta \over \delta \lambda(s)} \sum_{s'} ({\rm ln} \ Q)(s',s') \nonumber \\
&=& 2 \ Q^{-1}(s,s)
.\end{eqnarray}
and 
\begin{equation}
{\delta \over \delta \lambda(s)}Q^{-1}(s',s'') =2 \ Q^{-1}(s',s)Q^{-1}(s,s'') 
.\end{equation}
The stationary phase condition leads to 
\begin{equation}
1= \mbox{$\frac{3}{2}$}\biggl( {2 \over Q} \biggr)_{s,s}+
\int_0^L\!\int_0^L {\rm d}s' {\rm d}s'' {\bf f}(s')
Q^{-1}(s',s)Q^{-1}(s,s'') {\bf f}(s'') 
.\end{equation}
The stationary phase value $\lambda(s)$ thus depends on the value of
the external force ${\bf f}(s)$.  
The simplest but nontrivial case corresponds to the case of ${\bf
f}(s)={\bf f}={\rm const}$.  It is the case of stretching of 
semiflexible chains under constant value of ${\bf f}$ that is 
 appropriate to the recent experiments on DNA~\cite{SFB}.  
   If ${\bf f}$ is uniform,  then so
is $\lambda$.  In this case the stationary phase condition reduces to
\begin{eqnarray}
1&=&\mbox{$\frac{3}{2}$} \Biggl({1 \over- {l_{\rm p} \over 2} \bigl({\partial
\over \partial s'}\bigr)^2 +\lambda }\Biggr)_{s,s'}+{f^2 \over 4
\lambda^2} \nonumber \\
&=&\mbox{$\frac{3}{2}$} \sum_{-\infty}^{\infty} 
{ 1 \over \lambda L +{l_{\rm p} \over 2}{(2 \pi n)^2 \over L}}
+{f^2 \over 4 \lambda^2} 
.\end{eqnarray}
With the aid of Eq.~(\ref{ide}), this can be written in a closed form
\begin{equation}
1=\mbox{$\frac{3}{4}$} \sqrt{{2 \over l_{\rm p} \lambda}} \ 
 {\rm coth}(\mbox{$\frac{1}{2}$} \Omega L)
+{f^2 \over 4 \lambda^2}
\end{equation}
where $\Omega=\sqrt{2 \lambda \over l_{\rm p}}$.
As $L \rightarrow \infty$, this equation can be further simplified to yield
\begin{equation}
\label{spc1}
1-\mbox{$\frac{3}{4}$} \sqrt{2 \over l_{\rm p}
\lambda}={f^2 \over 4 \lambda^2}
.\end{equation}
As $f \rightarrow 0$, this equation reduces to the earlier one in
Eq.~(\ref{SPC}) as expected. \\

\vspace{0.2in}
\noindent
{\bf 4.2  Conformation of a semiflexible chain under tension} \\

In the case of uniform ${\bf f}$, we obtain a simple expression
for the correlation function:
\begin{eqnarray}
\bigl< {\bf u}(s) \cdot {\bf u}(s') \bigr>&=&3 Q^{-1}(s,s')+
{f^2 \over 4 \lambda^2} \nonumber \\
&=& \mbox{$\frac{3}{2}$}\sum_{-\infty}^{\infty} { {\rm exp} 
\bigl[i{2 \pi n \over L} |s'-s|
\bigr] \over \lambda L +{l_{\rm p} \over 2}{(2 \pi n)^2 \over L}}
+{f^2 \over 4 \lambda^2} \nonumber \\
&=& \mbox{$\frac{3}{4}$} \sqrt{{2 \over l_{\rm p} \lambda}} \ 
{{\rm cosh}[(L-2|s'-s|) \Omega/2] \over {\rm sinh}(\Omega L/2)}
+{f^2 \over 4 \lambda^2}
.\end{eqnarray}
where $\Omega=\sqrt{2 \lambda \over l_{\rm p}}$.  To derive the last step from 
the previous one, we used Eq.~(\ref{ide}).  As $L \rightarrow \infty$,
the above equation can be further simplified to yield
\begin{equation}
\label{corf}
\bigl< {\bf u}(s) \cdot {\bf u}(s') \bigr>=\mbox{$\frac{3}{4}$}
\sqrt{{2 \over l_{\rm p} \lambda}} \ {\rm e}^{-|s'-s| \Omega}
+{f^2 \over 4 \lambda^2}
.\end{equation}
If we set $|s'-s|$ to zero, this equation reduces to the stationary phase
condition.  Thus the free energy minimization condition in this case
also amounts to requiring ${\bf u}^2(s)=1$.  As ${\bf f} \rightarrow
0$, the coefficient in the exponential 
function becomes 1, resulting in the same expression as in Eq.~(\ref{cor1}).

Using the correlation function in Eq.~(\ref{corf}), the mean squared
internal distance of the chain under tension can be obtained
\begin{eqnarray}
\label{int.dis.f}
\bigl< |{\bf r}(s')-{\bf r}(s)|^2 \bigr>&=&\int_s^{s'}\!\int_s^{s'}
 \bigl< {\bf u}(s_1) \cdot {\bf u}(s_2) \bigr>{\rm d}s_1 {\rm d}s_2
\nonumber \\
&=&\mbox{$\frac{3}{4}$} \sqrt{{2 \over l_{\rm
p} \lambda}} 
\biggl[{2 \over \Omega}|s'-s|-{2 \over \Omega^2 }
(1-{\rm e}^{-\Omega |s'-s|} ) \biggr] 
+{ f^2 \over 4 \lambda^2}(s'-s)^2
.\end{eqnarray}
We can easily see that, for small $|s'-s|$, the first two term on the
R.H.S. of Eq.~(\ref{int.dis.f}) is dominant.  As $|s'-s|$ becomes larger,
the last term is more important.  At large length scales, the chain
conformation is thus mainly determined by the orienting field.  We can
now introduce a crossover 
length $S_{f}$ at which these two length scales coincide.  If the
crossover occurs at $S_f$ which is somewhat larger than
$\Omega^{-1}$, then we can get a simple expression for $S_{ f}$;
this can be obtained by
balancing the first term with the last one in R.H.S of 
Eq.~(\ref{int.dis.f})
\begin{equation}
S_{ f}=
\mbox{$\frac{3}{4}$} \sqrt{{2 \over l_{\rm p}
\lambda}} \biggl({4 \over f^2} {\lambda^2 \over
\Omega}\biggr) 
.\end{equation}
The condition, $S_{ f} \gg \Omega^{-1}$, is equivalent to 
${f^2 \over 4 \lambda^2} \ll {3 \over 4} \sqrt{{2 \over l_{\rm p}
\lambda}}= {\cal O}(1) $ or $f l_{\rm p} \ll 1$.  Under this
condition, the stationary 
phase condition coincides with that for an ideal semiflexible chain.  This
results in 
\begin{equation}
S_{f} \sim {1 \over f^2 l_{\rm p}}
.\end{equation}
The longitudinal size corresponding to $S_{ f}$ is called a {\it
tensile screening length},~\cite{Pincus} and is denoted by $\xi_{ f}$.  By noting 
that
parts of the chain within the length scale $S_{f}$ resemble an ideal
chain, we can have
\begin{equation}
\xi_{ f} \sim f^{-1}
.\end{equation}
The condition
${f^2 \over 4 \lambda^2} \ll 1$ for the above expression 
to be valid in turn amounts to requiring $f l_{\rm p} \ll 1$ or
equivalently $\xi_{ f} \gg l_{\rm p}$.  It is interesting to note
that results for $S_{f}$ and $\xi_{ f}$ given above are the
same as those for a Gaussian chain under tension.  This can be
understood as follows; 
 the condition, $\xi_{ f} \gg l_{\rm p}$,  
for the above results to be valid, is equivalent to saying that the
tensile screening length $\xi_{\rm f}$ contains large number of chain
segments of length $l_{\rm p}$.  The chain stiffness becomes marginal
in determining the shape of parts of chains within the length scale
$\xi_{f}$.  Since the effect of ${\bf f}$ is important only beyond this
length scale, the chain stiffness is not ``coupled'' to the external field
${\bf f}$. 

As ${f^2 \over 4 \lambda^2}$ becomes much larger than ${3 \over 4}
\sqrt{{2 \over l_{\rm p} \lambda}}$ or equivalently $f l_{\rm p}
\rightarrow \infty$ and thus $S_f$ approaches 0.  This implies
that, at any length scale beyond $S_f \approx 0$, the chain conformation
is governed by the interaction term $-\int {\bf f} \cdot {\bf u}(s)$.
As we will see below, as $l_{\rm p} f \rightarrow \infty$, the
entropical contribution is not balanced by the energy penalty for 
crumpled
conformation.  That is the chain segments are 
aligned with ${\bf f}$ as expected.  These quantitative 
considerations, which were already implicit in the calculations of 
 Fixman and Kovacs~\cite{F.K}, can be made quantitative using our general 
formalism. \\

\vspace{0.2in}
\noindent
{\bf 4.3  Elastic response of a semiflexible chain} \\

Let us now consider the average elongation $z$ defined by
\begin{equation}
\label{elong}
z=\biggl< \int_0^L {\rm d}s{\bf u}(s) \cdot {{\bf f} \over f}\biggr>
.\end{equation}
This quantity, which is experimentally measureable, shows how the chain 
responses to the tangential
field.  The statistical average in Eq.~(\ref{elong}) can be conveniently
expressed in  
terms of the free energy functional ${\cal F}$ (see Eq.~(\ref{free.f})).  For the case of
uniform ${\bf f}$, we obtain a simple expression for this:
\begin{eqnarray} 
z={{\bf f} \over f} \cdot {\delta {\cal F} \over \delta
{\bf f}}&=& f \int_0^L \biggl[ -l_{\rm p} \Bigl({\delta \over
\delta s }\Bigr)^{2}+2 \lambda \biggr]^{-1} {\rm d}s \nonumber \\
&=&  f {L \over 2 \lambda (f)}
.\end{eqnarray}
where the argument in $\lambda(f)$ is introduced to emphasize the
dependence of $\lambda$ on $f$.  This equation along with the
stationary phase condition Eq.~(\ref{spc1}) provide the average elongation with
respect to $f$.   To obtain analytic expressions, we first take two limiting
 cases
of $l_{\rm p} \rightarrow 0$ and $f \rightarrow \infty$.  As $l_{\rm p}
\rightarrow 0$ or more precisely $l_{\rm p}f \rightarrow 0$, the
problem reduces to that of a stretched Gaussian chain as mentioned in 
sec. 4.1.  
In this case, we can get the following elastic response relation
\begin{equation}
z=f {R_0^2 \over 3}
\end{equation}
where $R_0^2=2 l_0 L=2({2 \over 3} l_{\rm p}) L$ is the size of the
corresponding ideal semiflexible chain. 
This is the elastic response relation of a Gaussian chain under
tension~\cite{De}. 
For the case of $f l_{\rm p} \ll 1$, the stationary phase condition can be expanded   
in terms of $f^2$.  Up to $f^2$, we get
\begin{equation}
\lambda(h) \approx \lambda +{f^2 \over 2 \lambda}
\end{equation}
where $\lambda$ is the stationary phase condition for $f=0$.
Thus the average elongation is given by
\begin{equation}
\label{Rh0}
z \approx f {R_0^2 \over 3}[1-\mbox{$\frac{8}{9}$} f^2 l_0^2]
.\end{equation}
Note here that the control parameter in this expansion is $f l_{\rm p}$ (or $f l_0$).
  
As $l_{\rm p}$ increases the term ${f^2 \over 4 \lambda^2}$ in the stationary 
  phase condition becomes important.  If the value of $l_{\rm p}$ reaches the same order of magnitude as that
  of $\xi_f$, we expect the effect of the orienting field to be 
important over all length scales.  In other words, chain fluctuations become 
frozen 
 as $l_{\rm p} f$ becomes large.   As $f \rightarrow \infty$, the  stationary phase
 condition has a solution at $ \lambda^{-1}=0$, resulting in $\lambda=
 {1 \over 2}f$.
The average elongation is thus given by $z=L$.  Chain fluctuation are totally
suppressed in this case.  For finite but much larger $f$ than $l_{\rm p}$, 
$\lambda(f)$ can be approximated as 
\begin{equation}
\lambda(f)=
\mbox{$\frac{1}{2}$}f 
\Bigl(1+ \mbox{$\frac{3}{4}$} \sqrt{1\over l_{\rm p}f } \Bigr)
.\end{equation}
Accordingly $z$ for finite $f$ is smaller than $L$, which is due to chain 
fluctuations. 
\begin{equation}
\label{R_f}
z \approx  L \Bigl(1-\mbox{$\frac{3}{4}$} \sqrt{1\over l_{\rm p}f } \Bigr)
.\end{equation}
Similar result with slightly larger numerical factor was obtained by
T. Odijk~\cite{21.} considering a semiflexible chain near the rod
limit.

Since the chain under strong elongation as described by Eq.~(\ref{R_f}) 
is
nearly parallel to the orienting force ${\bf f}$, we can approximate
$z$ as 
\begin{eqnarray}
\label{R_h1}
z&=&\biggl< \int_0^L {\bf u}(s) {\rm d}s \cdot {{\bf f} \over f} \biggr> 
\nonumber \\
&=&\biggl<\int_0^L {\rm cos} \ \theta_f(s) {\rm d} s \biggr> \nonumber \\
&=&L \Bigl[1-\mbox{$\frac{1}{2}$} \bigl< \theta_f^2 (s) \bigr> \Bigr]
\end{eqnarray}
where $\theta(s)$ is the angle between ${\bf f}$ and ${\bf u}(s)$.
For a uniform field, we expect $\bigl<\theta_f^2(s) \bigr>$ not to depend on $s$
for a long chain, i.e., $\bigl<\theta_f^2(s) \bigr>=\bigl<\theta_f^2 \bigr>$.  
Comparing
Eq.~(\ref{R_f}) and~(\ref{R_h1}), we can obtain the expression for this quantity:
\begin{equation}
\label{theta^2}
\bigl<\theta_f^2 \bigr> = \mbox{$\frac{3}{2}$} \sqrt{ {1 \over l_{\rm
p} f}}
.\end{equation}
This result again differs by numerical factor from the corresponding
expression for $\bigl<\theta_f^2 \bigr>$ derived by Odijk~\cite{21.}.

In the above analysis, we have seen that the stationary phase approach
reproduces results consistent with literature in limiting cases.  To
test our theory further, we solved the stationary phase condition
numerically.  To compare with experimental data obtained by Smith et
al.~\cite{SFB} on the stretching of DNA (see also Fig. 2 in Ref.~\cite{MS}) $l_0$ and $L$ have been
chosen to be $53.4 {\rm nm}$ and $32.80 \mu 
{\rm m}$ respectively.  In Fig. 2,
$z/L$ is 
plotted against ${\rm ln} \ f$.  Our numerical solution (continuous
curve) is well in quantitative agreement 
with the experimental data~\cite{SFB}.

In the previous analysis, we have seen that a strong orienting field tends
to suppress chain fluctuations as is implied by the correlation and
the elastic response relation.  This is also manifested in the
ratio ${\Delta R^2 \over <R^2>}$.  Using Eq.~(\ref{DeltaR^2}), we get
\begin{equation}
\Delta R^2 = \mbox{$\frac{3}{4}$} \sqrt{{2 \over l_{\rm p} \lambda}}
\biggl[ {2 \over \Omega} L-{2 \over \Omega^2}(1-{\rm e}^{-\Omega L})
\biggr]
.\end{equation}
and thus
\begin{equation}
{\Delta R^2 \over \bigl< R^2 \bigr>}={ \Delta R^2
\over \Delta R^2 +  {f^2 L^2 \over 4 \lambda^2}}
.\end{equation} 
For a short chain, ${\Delta R^2 \over <R^2>}={\cal O}(1)$ while it
approaches zero as $L \rightarrow \infty$.  It can be easily seen
that value of this quantity crosses over at the contour length $L
\approx S_f$ from fluctuation dominated limit (${\Delta R^2 \over
<R^2 >}={\cal O}(1)$) to fluctuation suppressed limit  (${\Delta R^2
\over <R^2 >}\approx 0$).  As $l_{\rm p} f \rightarrow \infty$,
$S_f$ approaches zero.  Thus in this case, chain fluctuation
becomes totally suppressed for a chain with contour length $L \gg S_f
\approx 0$.   \\

\vspace{0.2in}
\noindent
{\bf 4.4  Operator representation} \\

Just as in the case of an ideal semiflexible chain, it is 
instructive to use
operator representation of the free energy functional ${\cal
F}[\lambda,{\bf f}]$.  The corresponding Hamiltonian in this case
describes a quantum harmonic oscillator in the linear force field
described by ${\bf f}$:
\begin{eqnarray}
\hat{\cal H}[\lambda,{\bf f}]&=&\hat{\cal H}[\lambda]+\Delta \hat{\cal
H} \nonumber \\
&\equiv&{\hat{\bf p}^2 \over 2 l_{\rm p}}+\lambda \hat{\bf u}^2-\hat{\bf u}
\cdot{\bf f} \nonumber \\
&=&{\hat{\bf p}^2 \over 2 l_{\rm p}}+\lambda \biggl(\hat{\bf u}-{{\bf
f} \over 2 \lambda} \biggr)^2-{f^2 \over 4 \lambda}
.\end{eqnarray}
Since $[\hat{\bf p},{\bf f}]=0$ and thus the commutator
$[\hat{p_j},\hat{u_k}]=-i \delta_{jk}$ is invariant under $\hat{\bf u}
\rightarrow \hat{\bf u}-{{\bf f} \over 2 \lambda}$, the external field
just shifts eigenvalues of $\hat{\cal H}[\lambda,{\bf f}]$ by the same
amount $-{f^2 \over 4 \lambda}$.  This results in
\begin{equation}
{\rm tr} \ {\rm e}^{-L \hat{\cal H}[\lambda,{\bf f}]}=\biggl({1 \over
2 \ {\rm sinh} (\mbox{$\frac{1}{2}$}\Omega L)} \biggr)^3 \ {\rm exp}\biggl( 
{f^2 \over 4 \lambda}L \biggr)
.\end{equation}
This leads to the same stationary phase condition as in Eq.~(\ref{spc1}).  

Correlation function of ${\bf u}(s)$ can be also computed following
the similar steps to that leading to Eq.~(\ref{cor1})
\begin{eqnarray}
\bigl< {\bf u}(s) \cdot {\bf u}(s') \bigr>&=&\sum_{\bf n} 
\Bigl|
\bigl<0| \Bigl( \hat{\bf u}+{{\bf f} \over 2 \lambda} \Bigr) |
{\bf n} \bigr> \Bigr|^2 \ {\rm e}^{-(E_{\bf n}-E_0)|s'-s|} \nonumber
\\
&=&|
\bigl<0| \hat{\bf u} |
{\bf n} \bigr> |^2 \ {\rm e}^{-(E_1-E_0)|s'-s|}
+{f^2 \over 4 \lambda^2} \nonumber \\
&=&\mbox{$\frac{3}{4}$} \sqrt{{2 \over l_{\rm p} \lambda}} \ {\rm
e}^{-|s'-s| \Omega}+{f^2 \over 4 \lambda^2}
.\end{eqnarray}
The operator representation again produces the same result as the
functional integral formalism.  One of the advantages of using the operator
representation is that we can use ground state dominance approximation
provided there is a energy gap.  \\

\vspace{0.5in}
\noindent
{\bf 5.  SEMIFLEXIBLE CHAINS IN A NEMATIC FIELD} \\

As a final example of the utility of our approach we consider an
semiflexible chain in a nematic environment.   These calculations 
expose conditions under which the stationary phase approach is not successful.  
The interaction energy of a chain in a
nematic field along ${\bf n}$ is assumed to be $-g \int_0^L [{\bf u}(s)
\cdot {\bf n} ]^2$, where the coupling constant $g$ measures the
strength of nematic field.   Thus conformations of nematic polymers
either parallel or 
anti-parallel to the nematic field are equally energetically
favorable.  This is in contrast to the case of a stretched chain.
The Hamiltonian of a single chain in the present case is given by~\cite{7.,KDN}
\begin{equation}
{{\cal H}[g] \over k_{\rm B} T}=\int_0^L {\rm d}s \Biggl\{ {l_{\rm p} \over 2}
\biggl({\partial {\bf u} \over \partial s} \biggr)^2 -g [{\bf u}(s)
\cdot {\bf n}]^2 \Biggr\}
\end{equation}
where ${\bf u}(s)$ is defined only on a unit sphere of $|{\bf
u}(s)|=1$.  At the meanfield level this takes the form
\begin{equation}
\label{H.N}
{{\cal H}[\lambda,g] \over k_{\rm B} T}=
\mbox{$\frac{1}{2}$} 
\int_0^L\!\int_0^L {\rm d}s {\rm d}s'\ {\bf u}(s) \ Q(s,s') \ {\bf
u}(s')-g \int_0^L {\rm d}s \ {\bf u}_z^2(s)
\end{equation}
such that the weight is given by
\begin{equation}
\Psi_{\rm MF}€[{\bf u}(s),g] \propto {\rm exp} \biggl\{-{{\cal H}[\lambda,g]
\over k_{\rm B}T } \biggr\}
\end{equation}
where the direction of the nematic field is chosen to be parallel to
$z$-axis.
With the convention ${\bf u}_\perp=(u_x,u_y,0)$, this can be rewritten
as 
\begin{equation}
{{\cal H}[{\bf u}(s)] \over k_{\rm B} T}=\mbox{$\frac{1}{2}$} 
\int_0^L\!\int_0^L {\rm d}s {\rm d}s' \Bigl[ {\bf u}_\perp (s) \ Q(s,s') \
{\bf u}_\perp (s')+u_z(s) \ Q_1(s,s') \ u_z(s') \Bigr]
\end{equation}
where
\begin{equation}
Q(s,s')=\biggl[-l_{\rm p} \biggl({\partial \over \partial s} \biggr)^2 +2
\lambda \biggr] \ \delta(s'-s)
\end{equation}
and
\begin{equation}
Q_1(s,s')=\biggl[-l_{\rm p} \biggl({\partial \over \partial s}\biggr)^2+2
(\lambda-g) 
\biggr] \ \delta(s'-s)
.\end{equation}
The corresponding generating functional ${\cal F}[\lambda,g,{\bf
f}(s)]$, up to an additive constant, now has the following form
\begin{eqnarray}
{\cal F}[\lambda,g]\!\!\!&=&\!\!\!{\rm tr \ ln} \ Q-\mbox{$\frac{1}{2}$}\int_0^L\!\int_0^L
{\rm d}s {\rm d}s'{\bf f}_\perp (s) Q^{-1}(s,s') {\bf f}_\perp (s')
\nonumber \\
&+&\!\!\!\mbox{$\frac{1}{2}$}{\rm tr \
ln} \ Q_1-\mbox{$\frac{1}{2}$}\int_0^L\!\int_0^L 
{\rm d}s {\rm d}s'{ f}_z (s)  Q_1^{-1}(s,s')  { f}_z (s') -\int_0^L 
\lambda(s) {\rm d}s
\end{eqnarray}
from which we can obtain
\begin{eqnarray}
\bigl< {\bf u}_\perp (s) \cdot {\bf u}_\perp (s') \bigr>&=&-{\delta \over
\delta {\bf f}_\perp (s)} \cdot {\delta {\cal F} \over \delta {\bf
f}_\perp (s')}+\bigl< {\bf u}_\perp (s) \bigr> \cdot \bigl< {\bf
u}_\perp (s') \bigr> \nonumber \\
&=& \mbox{$\frac{1}{2}$}
\sqrt{{2 \over l_{\rm p} \lambda}} \ {\rm e}^{-|s'-s| \Omega}
\end{eqnarray}
and
\begin{eqnarray}
\bigl< {u}_z (s) \cdot { u}_z (s') \bigr>&=&-{\delta \over
\delta {f}_z (s)} \cdot {\delta {\cal F} \over \delta {
f}_z (s')}+\bigl< { u}_z (s) \bigr> \cdot \bigl< {
u}_z (s') \bigr> \nonumber \\
&=& \mbox{$\frac{1}{4}$}
\sqrt{{2 \over l_{\rm p} (\lambda-g)}} \ {\rm e}^{-|s'-s| \Omega_1}
\end{eqnarray}
with $\Omega=\sqrt{2 \lambda \over l_{\rm p}}$ and $\Omega_1=\sqrt{2
(\lambda-g) \over l_{\rm p}}$.   
In the last steps of above equations, we set ${\bf f}(s)$ to zero.
The stationary phase value of $\lambda$ now satisfies the following equality
\begin{equation}
\label{SPC.N}
1=\mbox{$\frac{1}{2}$} \sqrt{2 \over l_{\rm p} \lambda}+\mbox{$\frac{1}{4}$} 
 \sqrt{2 \over l_{\rm p} (\lambda-g)}
\end{equation}
which ensures
\begin{equation}
1=\bigl<{\bf u}_\perp^2(s) \bigr> + \bigl< u_z(s) \bigr> =\bigl< {\bf
u}^2(s) \bigr>
.\end{equation}

It can be easily seen that the stationary phase value of $\lambda$ in
this case is larger than that for the case of free chains.  But
$(\lambda-g)$ is smaller than that for the latter case.  This implies
that the nematic field 
induces ordering along the field but decrease the persistence length perpendicular to it.  That is
\begin{eqnarray}
\bigl< R_\perp^2 \bigr>&=&\int_0^L\!\int_0^L \bigl<
{\bf u}_\perp (s) \cdot
{\bf u}_\perp(s') \bigr> {\rm d}s {\rm d}s' \nonumber \\
&=&\mbox{$\frac{2}{3}$} (2 l_{\rm eff}^\perp L) \nonumber \\
&<& 
\mbox{$\frac{2}{3}$} (2 l_0 L)
\end{eqnarray}
and 
\begin{eqnarray}
\bigl< R_z^2 \bigr>&=&\int_0^L\!\int_0^L \bigl< { u}_z (s) \cdot
{ u}_z(s') \bigr> {\rm d}s {\rm d}s' \nonumber \\
&=&\mbox{$\frac{2}{3}$} (2 l_{\rm eff}^z L) \nonumber \\
&<& 
\mbox{$\frac{1}{3}$} (2 l_0 L)
\end{eqnarray}
where $l_0= \mbox{$\frac{2}{3}$}l_{\rm p}$ is the persistence length
of a free chain.  The size of chains in nematic field does
not grow as $L$ as $L \rightarrow \infty$, as is the case for the
stretched chain within the meanfield variational theory.  To be more precise, let us consider the stationary
phase solutions for two extreme cases.  In the following derivations any quantity with (without) argument $g$
corresponds to a nematic polymer (non-interacting polymer). 

\noindent
(a) {\it The weak nematic limit}, $g \rightarrow 0$. 

For small $g$, we can write $\lambda(g) \approx \lambda (1+
\epsilon)$.  By assuming both $\epsilon$ and $g$ small, we can have
\begin{equation}
\lambda(g) \approx \lambda (1+\mbox{$\frac{2}{3}$} g l_0)
\end{equation}
and thus
\begin{equation}
\Omega(g)=\sqrt{2 \lambda(g)/ l_{\rm p}} \approx 
          \sqrt{2 \lambda / l_{\rm p}} \ (1+\mbox{$\frac{2}{3}$}g
l_0)
\end{equation}
and
\begin{equation}
\Omega'(g)=\sqrt{2 (\lambda(g)-g) / l_{\rm p}} \approx 
          \sqrt{2 \lambda / l_{\rm p}} \ (1-\mbox{$\frac{1}{3}$}g
l_0) 
.\end{equation}
These lead to 
\begin{equation}
{\bigl< R_\perp^2 \bigr> \over \mbox{$\frac{2}{3}$} ( 2 l_0 L)} \approx
1-\mbox{$\frac{1}{3}$}g l_0 
\end{equation}
and
\begin{equation}
{\bigl< R_z^2 \bigr> \over \mbox{$\frac{1}{3}$}( 2 l_0 L)} \approx
1+\mbox{$\frac{1}{3}$}g l_0 
.\end{equation}
Thus the persistence length along the nematic field is increased by
the factor of $(1+\mbox{$\frac{1}{3}$}gl_0)$ which can be compared with
$(1+\mbox{$\frac{2}{3}$}gl_0)$ (in our notation)
obtained by Warner at al.~\cite{7.}.

\noindent
(b) {\it The strong nematic limit}, $g \rightarrow \infty$.

If we simply let $g \rightarrow \infty$ in the stationarity condition in
Eq.~(\ref{SPC.N}), then $\lambda(g)$ and thus $\Omega(g)$ should
diverge so that 
$(\lambda(g)-g)$ remain positive.  This results in 
\begin{equation}
\bigl< {\bf u}_\perp (s) \cdot {\bf u}_\perp(s') \bigr>=0
\end{equation}
and
\begin{equation}
\bigl< {u}_z (s) \cdot { u}_z(s') \bigr>={\rm exp}(-|s'-s|/2 l_{\rm 
p})
.\end{equation} 
The diverging nematic field totally suppresses chain fluctuations 
perpendicular to it but increases the persistence length parallel to
it by the factor of 2.   The increase of the persistence length along
the nematic field by only numerical factor, however, is {\it not} correct. 
Several authors predicted an effective persistence which grows exponentially
with $\sqrt{g l_{\rm p}}$~\cite{7.,VO,KDN,Genn}.  Following the
reference~\cite{KDN},  this can be explained as follows; since the
interaction energy $-g u_z^2$ which is quadratic in $u_z$ has two
minima for $u_z=\pm 1$, the ``tunneling'' probability from one minimum to
the next is very small for large $g$.  This results in the length scale
which varies exponentially as $\sqrt{g l_{\rm p}}$ over which the
chain is parallel to the nematic field.  For more details see the
reference~\cite{KDN}.  

The stationary phase free energy in Eq.~(\ref{H.N}) also contains the
quadratic interaction term $-g u_z^2$.  This term if added to $\lambda
u_z^2$ becomes $(\lambda-g) u_z^2$ with positive $(\lambda-g)$ which
has only {\it one } minimum at $u_z=0$.  Thus the stationary phase
approach does not account for the instanton solution,
resulting in much smaller persistence length than the exact results.   This 
problem illustrates the possible limitations of the
stationary phase approach; if there is a possibility of broken 
symmetry in the problem then the meanfield variational theory should 
be used with caution.  \\

\vspace{0.5in}
\noindent
{\bf 6.  CONCLUSIONS} \\

In this chapter we have described a simple meanfield variational
approach to study a number of properties of intrinsically stiff
chains which are appropriate models for a large class of biopolymers.
The exact statistical mechanics of such systems are complicated by
the fact that the constraint of the tangent vector being unity has to
be enforced at all points along the contour. In the meanfield approach
this local constraint, which is difficult to impose, is replaced
by a global one. The global constraint ensures that the condition
of the tangent vector being unity is obeyed on an average. We have
described the calculation of the distribution of end-to-end distance and
the elastic response of stiff chains under tension using this basic
methodology. In the former example we find that the simple expression
almost quantitatively fits the results of simulation. For the case
of the stiff chain under tension we recover analytically all the known limits.
Furthermore we obtain quantitative agreement with recent experiments on the
stretching of DNA using the total contour length and the persistence
length as adjustable parameters. The limitations of our approach become
obvious in situations that involves broken symmetry such as the case of
a stiff chain in a strong nematic potential. Nevertheless it is clear
that the meanfield variational approach lays the foundation for
systematic studies of more complicated problems involving semiflexible
chains.


\newpage
\begin{thebibliography}{72} 
\bibitem{1.}See, for example, P. J. Flory, {\it Statistical Mechanics of Chain Molecules} (Interscience,
 New York, 1969); K. F. Freed, Adv. Chem. Phys. {\bf 22}, 1 (1972);
 M. Doi and S. F. Edwards, {\it The Theory of Polymer Dynamics} (Oxford   
 University Press, Oxford, 1986).  
\bibitem{2.}See, for example, J. des Cloiaeaux and G. Jannink, {\it Polymers
in Solution, Their Modelling and Structure} (Oxford University Press, Oxford,
1990); M. Doi and S. F. Edwards, {\it The Theory of Polymer Dynamics} (Oxford
University Press, Oxford, 1986).
\bibitem{3.} H. Yamakawa, {\it Modern Theory of Polymer Solutions} (New York: 
            Harper and Row, 1971). 
\bibitem{Frey}  K. Kroy and E. Frey, Phys. Rev. Lett. {\bf 77}, 
306 (1996) and references therein.  
\bibitem{SFB} S. B. Smith, L. Finzi, C. Bustamante, Science {\bf 258},
1122 (1992).

\bibitem{4.} N. Saito, J. Takahashi, and Y. Yunoki, J. Phys. Soc. Jpn. 
           {\bf 22}, 219 (1967).
\bibitem{5.}R. A. Harris and J. E. Hearst, J. Chem. Phys. {\bf 44},
2595 (1966).
\bibitem{6.} K. F. Freed, J. Chem. Phys. {\bf 54}, 1453 (1971); M. G. Bawendi
and K. F. Freed, J. Chem. Phys. {\bf 83}, 2491 (1985).
\bibitem{7.} M. Warner, J. M. F. Gunn, and A. B. Baumgartner, 
J. Phys. A {\bf 18}, 3007 (1965); M. Warner, J. M. F. Gunn, and A. B. Baumgartner,
J. Phys. A {\bf 19}, 2215 (1986).    
\bibitem{8.} For a review of functional integral methods in polymers
 see K. F. Freed, Adv. Chem. Phys. {\bf 22}, 1 (1972).
\bibitem{Ts} V. N. Tsvetkov, {\it Rigid Chain Polymers} (Consultants
Bureau, New York, 1989).
\bibitem{VO} G. J. Vroege and T. Odijk, Macromolecules {\bf 21}, 2848
(1988).
\bibitem{}  T. Odijk, Macromolecules {\bf 16}, 1340 (1983);  T. Odijk,
Macromolecules {\bf 19}, 2313 (1986).

\bibitem{HT3} B.-Y. Ha and D. Thirumalai, J. Chem. Phys. {\bf 103} 9408 (1995).

\bibitem{9.} A. M. Gupta and S. F. Edwards, J. Chem. Phys. {\bf 98}, 1588 (1993);
           Z. Y. Chen, Phys. Rev. Lett. {\bf 71}, 93 (1993). 
\bibitem{HT2}B. -Y. Ha and D. Thirumalai, Macromolecules {\bf 28}, 577
(1995). 
\bibitem{BJ}  J. L. Barrat and J. F. Joanny, Europhys. Lett. {\bf 
24}, 333 (1993).        
           
\bibitem{12.}J. B. Lagowski, J. Noolandi and B. Nickel, J. Chem. Phys. {\bf 95}, 
1266 (1991).
\bibitem{14.} R. G. Winkler, P. Reineker, and L. Harnau, J. Chem. Phys. 
{\bf 101}, 8119 (1994). 
\bibitem{15.}K. Soda, J. Phys. Soc. Jpn. {\bf 35}, 866 (1973).
\bibitem{16.} D. J. Amit, {\it Field Theory, the Renormalization Group,  and 
      Critical Phenomena} (World Scientific, Singapore, 1978).
\bibitem{17.}R. P. Feynman and A. R. Hibbs, {\it Quantum Mechanics and   
     Path Integral} (McGraw-Hill, Inc., 1965).

 \bibitem{19.} O. G. Berg, Biopolymer {\bf 18}, 2861 (1979).
\bibitem{20.} K. Huber, W. H. Stockmayer, and K. Soda, Polymer {\bf 31}, 1811 (1990).
\bibitem{18.} J. Zinn-Justin, {\it Quantum Field Theory and Critical Phenomena}
(Oxford University Press, Oxford, 1989).
\bibitem{W.F}  J. Wilhelm and E. Frey, Phys. Rev. Lett. {\bf 77}, 
2581 (1996). 
\bibitem{otto} M.Otto, J. Eckert, and T. A. Vilgis, Macromol. Theory Simul. {\bf3}, 543 (1994).
\bibitem{F.K}  M. Fixman and J. Kovacs, J. Chem. Phys. {\bf 58}, 1564 
(1973). 
\bibitem{21.} T. Odijk, Macromolecules {\bf 28}, 7016 (1995).
\bibitem{HTInt} B.-Y. Ha and D. Thirumalai, J. Chem. Phys. {\bf 106}, 
4243 (1997). 
\bibitem{KV} A. Kholondenko and T. Vilgis, Phys. Rev. E {\bf 50},
1257 (1994). 
\bibitem{MS} J. F. Marko and E. D. Siggia, Science {\bf 265}, 1599
(1994).
\bibitem{MSInt} J. F. Marko and E. D. Siggia, Macromolecules {\bf 28}, 8759
(1995).
\bibitem{Pincus} P. Pincus, Macromolecules {\bf 9}, 386 (1976).
\bibitem{De} P.-G. de Gennes, {\it Scaling Concepts in Polymer 
Physics} (Cornell University Press, Ithaca and London, 1979).

\bibitem{KDN} K. D. Kamien, P. L. Doussal, and D. R. Nelson,
Phys. Rev. A {\bf 45}, 8727 (1992).
\bibitem{Genn} P. G. de Gennes, in {\it Polymer Liquid Crystals},
edited by A. Ciferri, W. R. Kringbaum, and R. B. Meyer (Academic
Press, New York, 1982).  
\end{thebibliography}
\end{document}